\documentclass[acmtog, authorversion]{acmart}

\usepackage{booktabs} % For formal tables
\usepackage{graphicx}
\usepackage{amsmath,bm,bbm,dsfont}
\usepackage{subfigure}
\setcitestyle{square}

\begin{document}
% Title portion
\title{Shape-from-Mask: A Deep Learning Based Human Body Shape Reconstruction from Binary Mask Images}

\author{Zhongping Ji}
\orcid{1234-5678-9012-3456}
\affiliation{%
  \institution{Hangzhou Dianzi University}
  \city{Hangzhou}
  \state{Zhejiang}
  \country{China}}
\email{jzp@hdu.edu.cn}
\author{Xiao Qi}
\affiliation{%
  \institution{Hangzhou Dianzi University}
  \city{Hangzhou}
  \state{Zhejiang}
  \country{China}}
\author{Yigang Wang}
\affiliation{%
  \institution{Hangzhou Dianzi University}
  \city{Hangzhou}
  \state{Zhejiang}
  \country{China}}  
\author{Gang Xu}
\affiliation{%
  \institution{Hangzhou Dianzi University}
  \city{Hangzhou}
  \state{Zhejiang}
  \country{China}}  
\author{Peng Du}
\affiliation{%
  \institution{Hangzhou Dianzi University}
  \city{Hangzhou}
  \state{Zhejiang}
  \country{China}}  
\author{Qing Wu}
\affiliation{%
  \institution{Hangzhou Dianzi University}
  \city{Hangzhou}
  \state{Zhejiang}
  \country{China}}  
%\email{beranger@inria.fr}

\renewcommand\shortauthors{Zhongping, Ji. et al}

\begin{abstract}
3D content creation is referred to as one of the most fundamental tasks of computer graphics.
And many 3D modeling algorithms from 2D images or curves have been developed over the past several decades.
Designers are allowed to align some conceptual images or sketch some suggestive curves, from front, side, and top views, 
and then use them as references in constructing a 3D model automatically or manually. 
However, to the best of our knowledge, no studies have investigated on 3D human body reconstruction in a similar manner.

In this paper, we propose a deep learning based reconstruction of 3D human body shape from 2D orthographic views.
A novel CNN-based regression network, with two branches corresponding to frontal and lateral views respectively, 
is designed for estimating 3D human body shape from 2D mask images.
We train our networks separately to decouple the feature descriptors which encode the body parameters from different views, 
and fuse them to estimate an accurate human body shape.
In addition, to overcome the shortage of training data required for this purpose, 
we propose some significantly data augmentation schemes for 3D human body shapes, 
which can be used to promote further research on this topic.
Extensive experimental results demonstrate that visually realistic and accurate reconstructions can be achieved effectively using our algorithm.
Requiring only binary mask images, our method can help users create their own digital avatars quickly, 
and also make it easy to create digital human body for 3D game, virtual reality, online fashion shopping.  
\end{abstract}





\maketitle

\section{Introduction}
The real-world data are three-dimensional (3D), 
but human beings have learned to recognize and reason about 3D objects based upon their two-dimensional (2D) appearances from various viewpoints.
In computer vision community,
amounts of algorithms based on feature descriptors of 2D images are developed to classify, detect and recognize new 2D photographs of those underlying 3D objects.
Recently, deep learning accelerates a huge boost to the already rapidly developing field of computer vision. 
With deep learning, a large number of new applications of computer vision techniques have been introduced and are now becoming parts of our everyday lives.
This booming progress is mainly benefited from the convolutional neural networks (CNNs) which designed for 2D images with a grid-structure. CNNs demonstrate ramarkable flexibility and performance in a wide range of computer vision tasks.
These tasks can be achieved by learning some regression models from huge image data with lables,
without the prior knowledge of underlying 3D objects.
The well trained deep CNNs have the ability to automatically extract useful features of 2D images and to `remember' the correspondences between the extracted features and the lables. 
However, another fundamental task of computer vision, to draw inferences about entire 3D objects from their 2D images, is still a challenge. 
The CNNs seem as an inappropriate candidate for this problem.
One of the reasons is that they require a huge amount of real-world training data which are inpractical to acquire for general 3D objects.
Recently, CNNs are trained to estimate specific type of 3D shapes, such as 3D faces, from 2D images.
Richardson et al. synthesize 3D faces by sampling from the 3D morphable model (3DMM) and then use a CNN to regress parameters of 3DMM for 3D face estimating \cite{RichardsonSK16}.
Tu\~{a}n Tr\~{a}n et al. train a CNN to regress parameters of a 3D morphable shape and texture model directly from an input photo \cite{tran2016regressing}.
To overcome the shortage of training data for a very deep CNN, they propose a pooling method to synthesize huge numbers of faces and their 3D morphable model representations. 
These work demonstrates that CNNs are capable of reconstructing 3D shapes from 2D images for specific type of objects.

In a parallel research field, computer graphic, content creation in 3D space is referred to as one of the most fundamental tasks. 
The workflow of traditional 3D modeling on a computer can be described in two phases. 
A designer draws some 2D suggestive curves or sketches for a 3D object, usually from three views (front, side, and top views), 
and then uses them as a reference in constructing a 3D shape. 
While the process of converting those 2D curves into a 3D shape is nontrivial, and often needs skills and time. 
The 3D position for every element of the shape must be specified manually to match the 2D designs. 
In computer graphic, translating 2D curves into 3D shapes automatically is a longstanding research topic.
The ultimate goal of 3D modeling is to allow users to convert a 2D design concept into a digital 3D model quickly.
Igarashi et al. present a modeling system for 3D freeform shape from 2D silhouettes interactive drawn by the user \cite{EVL1999}.
Nealen et al. develop a system for designing 3D freeform shapes with a set of curves \cite{NealenISA07}.
Rivers et al. build a system for translating 2D lines of an object from orthographic views into a 3D shape \cite{Rivers2010}.
The user focuses on designing the silhouettes in two or more orthographic views, 
and a 3D shape is automatically generated to matches those 2D cues. 
However, these traditional methods explicitly construct the translation from 2D curves to 3D shapes.
Building data-driven translations with a minimal amount of labor is still a major challenge.
Recently, CNNs have been designed to learn parameters from sketches for procedural modeling of man-made objects \cite{HuangKYM17}, buildings \cite{NGGBB16}, 3D face and caricature modeling \cite{HanGY17}. These methods demonstrate the effectiveness of CNNs for 3D modeling from 2D curves, and imply the 3D shapes might be highly coorelated with their 2D contours in image views. 

Inspired by these recent work on 3D modeling from 2D images and 2D curves, 
we attempt to reconstruct 3D human body shape from its 2D sections, using CNN-based networks.
However, automatic reconstruction human body from color images in clothing might lose the geometry accuracy due to the occlusions, 
whereas digital nude photographs will cause the privacy concerns.
Therefor, we intend to use only binary mask images to overcome these limitations.
The binary mask images not only include the 2D curves on silhouettes which can be extracted by the filters of CNN on the fly, 
but also imply the feature descriptors for local and global shapes (e.g. the body proportion inferred from regional areas) captured from a certain view.
In addition, 3D human body modeling is a complex task in computer graphics.
Actually, in many traditional 3D modeling packages, the human body modeling is always considered as a coarse-to-fine procedure. 
First, a coarse base mesh is created to represent the rough body shape.
Then it is subdivided into higher resolution, and added finer scale of geometric and texture details manually. 
Here we focus on reconstruct human body shapes with medium level of details which can be applied to many applications directly or imported into the some 3D modeling packages (e.g. ZBrush, Mudbox) for further processing if necessary.

In summary, this paper makes the following contributions,
% enumerate
\begin{enumerate}
\item Some significantly data augmentation schemes for 3D body shapes are introduced.
\item A novel CNN based deep regression network is designed for reconstructing 3D body shape from binary mask images directly, avoiding the need for extra annotations (e.g. segmentation and skeleton) and two stage training.
      \begin{enumerate}
      \item We design a two-branch network which represents features of different masks using different descriptors respectively.
      \item We use a DenseNet-like architecture to aggregate features of different scales to reduce the number of model parameters without losing effectiveness.
      \end{enumerate}
\item The proposed deep learning based 3D modeling scheme from two views may provide inspiration to 3D modeling from 2D images for other shapes, such as 3D faces/heads modeling.      
\end{enumerate}

\section{Related Work}
An intensive survey on human body shape modeling is beyond the scope of this paper. 
Technically we most inspired by the following two related areas: 
the space of human body shapes and the convolutional neural networks.
In this section, we review those most related work briefly.

\subsection{Data-driven human body space} 
Building statistical shape spaces for human bodies is a challenging task. 
The seminal work \cite{Allen2003} uses PCA to represent a space of human body shapes without considering the postures.
They make improvements to address hole-filling and detail from the template surface. 
The following work can be classified into two group of approaches.
Most methods learn shape-variation and posture-variation separately and combine them afterwards.
The representive method SCAPE \cite{Anguelov2005} builds a human shape model that spans variation in both shape and posture.
They learn a shape model from variations of different individuals and a posture model from a single subject.
This method and most following work encode the variations of human bodies in the terms of triangle deformation.
Jain et al. propose a simplified version of SCAPE to encode the variation in the terms of vertex coordinates,
and to model the postures with a skeleton-based surface skinning method \cite{JainTST10}.
Based on similar representation, Pishchulin et al. build a statistical body space from a large commercially available scan database and release the resulting model to research community \cite{PishchulinWHTS15}.
Another successful shape model SMPL is designed to represent shape-and-posture variations in vertex space \cite{LoperM0PB15}.
They encode the variations using an artist created mesh with a clean quad structure, which benefits some applications.
Recently, this model is used to estimate the posture and shape from a single image or a video \cite{Bogo2016,KBJM17,Varol0MMBLS17,pavlakos2018humanshape}.
Additional, another methods perform simultaneous analysis on shape and posture variations \cite{SCA06,HaslerSSRS09}.
These methods use a single model to explore both shape and posture variations, which can be used to perform realistic muscle deformation.

\subsection{Learning posture and shape from images}
Estimating 3D human shape and posture from multi-view images and RGB-D data is a promising topic and a remarkable progress has been made \cite{VlasicPBDPRM09,TongZLPY12,LiVGLBG13}.
However, these specialized sensors are not widely available which limits their extensive applications.
The estimation of 3D human posture and shape from a single monocular image is a cheaper scheme with many applications. 

The automatic reconstruction is a very challenging problem due to the strong ambiguities of estimating human body articulations from a single image.
The ambiguity is partially caused by self-occlusion, in clothing, large degrees of freedom for poses and the deformation of an articulated body. 
The early approaches focus on estimating an instance of SCAPE model that is consistent with 2D image
observations, like silhouettes, edges, shading, or 2D keypoints \cite{SigalNIPS2008,GuanWBB09}.
Their approaches rely on manually extracted silhouettes or some manual correspondence points. 
More recent works attempt to automate this effort due to the progress of deep learning.
Bogo et al. intend to solve this problem by exploiting the information of 3D body shape carried by 2D joints \cite{Bogo2016}.
They combine bottom-up and top-down strategies to estimate the pose and shape simultaneously. 
A CNN-based method, DeepCut \cite{pishchulin16cvpr}, is utilized to predict the 2D joints locations and then the SMPL model is fitted to the predicted 2D joints. 
And the reconstructed 3D model is also used to prevent the interpenetration. 
A regression tree model that predicts the 3D body parameters from 2D keypoints directly is presented \cite{LassnerUP2017}. 
Lassner et al. train a detector of $31$ segments and $91$ keypoints and then optimize the SMPL model parameters to estimate the 3D human posture and shape.
Pavlakos et al. predict some 2D heatmaps for joints and a silhouette for shape from a single
color image, and then train two networks to estimate the parameters of the human body posture and shape \cite{pavlakos2018humanshape}.
They demonstrate that the parameters of SMPL model can be predicted reliably from 2D keypoints and silhouettes.
These methods of recovering 3D human posture and shape focus on a multi-stage approach which combines the 2D joints, silhouettes or segmentations. 
Most of these methods rely on labor intensive tasks and interactive annotation tools to create the involved annotations or labels.
Varol et al. learn a fully convolutional network from a synthetic dataset of rendered SMPL bodies
to estimate the depth and body segmentation \cite{Varol0MMBLS17}.
Kanazawa et al. present an end-to-end framework for estimating a 3D human body from a
single monocular RGB image \cite{KBJM17}.
Their method infers 3D pose and shape parameters directly from image pixels,
training K+2 discriminator networks to judge the estimated SMPL parameters.
They demonstrate competitive results on tasks such as 3D joint location estimation and part segmentation.

The crucial part of these above mentioned methods is to estimate the posture by leveraging the relation between 3D joints and their 2D annoturations. 
The estimated shape partially acts as a regularization term to improve the accuracy of the estimated posture.
However, the accuracy of 3D shape is hard to evaluate due to the self-occlusion or in clothing.
Another weakness is that these methods heavily rely on the accuracy of 2D annotations to some extent.
Whereas, these work succeeds in building a bridge between 2D annotations and 3D shapes and demonstrates promising results which may benefit many applications.
From another point of view, our work focuses on training a model to estimate a human body shape only from binary mask images, without relying on the colors of pixels or additional annotations (such as joints, segmentations). 
The mask images can be derived from 3D models, grayscale/color images, or started from scratch via sketching or brushing.

\subsection{Convolutional neural networks}
Although CNN was originally introduced nearly three decades ago \cite{lecun89backpropagation}, only recently, 
it has enabled training of very deep layers to learn general purpose image descriptors in many computer vision tasks, 
such as image classification, object detection and semantic image segmentation.
As CNN becomes increasingly deep, the vanishing gradient issue emerges.
Highway networks \cite{Srivastava2015} and Residual networks \cite{HeZRS15} are designed to avoid the problem of vanishing gradient and to make training of very deep networks be feasible, bypassing signal between layers via identity connections. 
They share a critical characteristic: creating short paths from early layers to later layers.
Based on this insight, Huang et al. design a new CNN architecture, called DenseNet \cite{HuangLMW17}.  
DenseNet is composed of several dense blocks each of which is an iterative concatenation of previous feature maps. 
Each layer outputs $k$ features, where $k$ specifies the growth rate of the network. 
This architecture improves the efficiency in the parameter usage and makes a reduction in the number of parameters, 
performs iterative summation of previous layers to reuse the preceding feature maps,
and performs multi-scale supervision due to the short paths to all feature maps.  
These characteristics make DenseNets very competent for image classification \cite{HuangLMW17}, semantic segmentation \cite{JegouDVRB16}and image super-resolution \cite{iccvLLG17}.
Based on DenseNet, our architecture is designed to learn a feature representation that aggregates information from defferent layers. 
We show that these deep architectures can be adapted to regress parameters of a 3D human body model from binary mask images effectively.

\section{Body Shape Space}
We intend to build a PCA model that represents each body shape with a parameter vector $\varphi_s \in \mathbb{R}^k$
\begin{equation}{\label{eqn:pca}}
B_s=\bar B + \Omega \varphi_s
\end{equation}
where $B_s \in \mathbb{R}^{3N}$ is a body shape with $N$ vertices and the same topology, and $\Omega \in \mathbb{R}^{3N \times k}$ is composed of $k$ principal vectors $\{\omega_0,\omega_1,...,\omega_{k-1}\}$. Associated with each principal vector $\omega_i$ is a variance $\sigma_i^2$, and $\omega_i$ are sorted by $\sigma_i^2$ in descending order. $\bar B$ represents the mean body shape of the training dataset. This paper focuses on the learning and modeling of human body in a standard pose, thus our body shape space mainly covers variations in body shape and slight variations in the posture. 
As mentioned above, several body shape spaces have been built in the literature.
For the reality of the reconstructed body shape, it is quite important to model the variation of human shape using real-world data. 
Thus, our PCA model is built on 4308 body shapes released by \cite{PishchulinWHTS15}. 
Their work utilized CAESAR database \cite{Robinette99} which contains 3D scans of over 4500 American and European subjects in a standard pose. 
We uniformly normalize the human models into $[-1,1]$ for the $z-$coordinates and then rebuild our PCA model on the male and female data together.
To encode a large range of variations, we set $k=50$ in our implementation
which performs well in our extensive experiments.

\subsection{Data Augmentation}

Generally speaking, deep networks need massive amounts of training data to work well. 
However, existing 3D body shape datasets of real-world data are relatively small for this purpose, and acquiring large amounts of high-quality scans of 3D human body is labor intensive, expensive and impractical.
We therefore instead enlarge the dataset by leveraging the existing real-world data.

To increase the shape diversity, we perform data augmentation in three different manners, 
including sampling from the original data distribution, building segmented PCA models by segment-and-merge operations, and utilizing a generative adversarial network.

$\blacksquare$ 
\textbf{Interpolating and random sampling}.
Besides the 4308 models, we generate extra 5k models by linear interpolation.
These models are quite approximate to the original bodies, 
thus they are used to improve the proportion of original data in our final dataset.
Instead of randomly interpolating, we first obtain 100 classes by using a k-means clustering algorithm, and then interpolate the different cluster centroids to form new samples.
Further more, by sampling from the Gaussian distribution that the PCA
represents, we can generate an unlimited number of new individuals.
Specifically, we can presume that the coefficient $\varphi_s$ of PCA model in Eq. (\ref{eqn:pca}) is subject to a multivariate Gaussian distribution which described by the variances $\{\sigma_i^2\}$, thus we can randomly synthesize individuals $\varphi_s$ according to the variances $\{\sigma_i^2\}$.
Finally, we generate extra 20k models to expand our dataset.

\begin{figure}
  \centering
  \subfigure[]{
    \label{fig:mixture2:a} 
    \includegraphics[angle=0,width=1.44in]{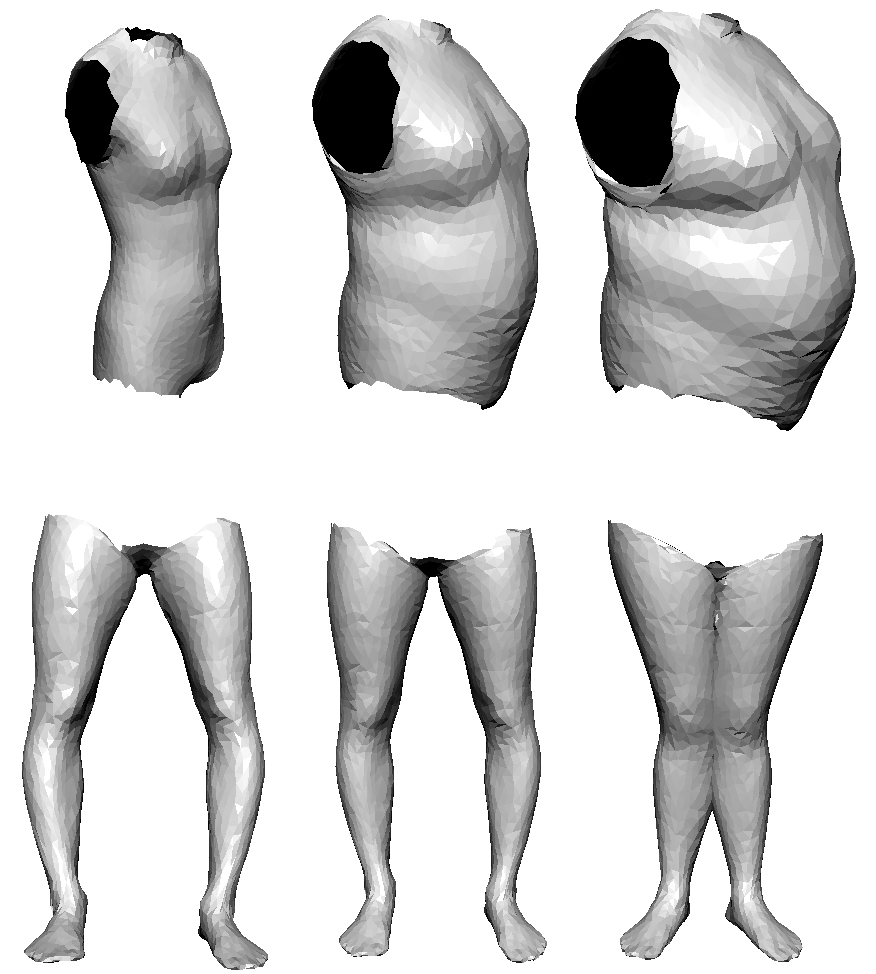}}
  \hspace{-0.08in}
  \subfigure[]{
    \label{fig:mixture2:b} 
    \includegraphics[angle=0,width=0.85in]{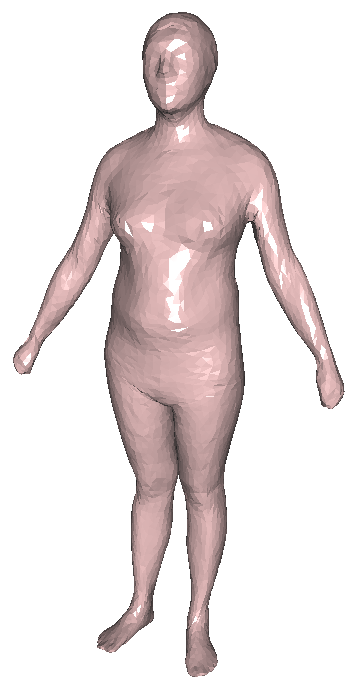}}
    \hspace{-0.08in}
  \subfigure[]{
    \label{fig:mixture2:c} 
    \includegraphics[angle=0,width=0.85in]{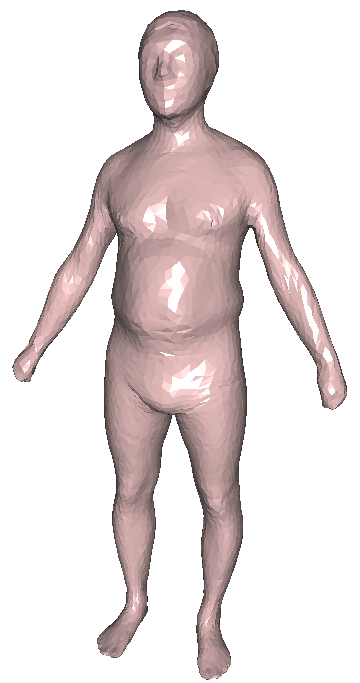}}
  \caption{\label{fig:mixture2}
Some representive components and reconstructions are shown.
  (a) Different types of torsos and legs (O-type legs and X-type legs);
  (b-c) two individuals created by merging a torso with different types of legs.
}
\end{figure}

$\blacksquare$
\textbf{Segmented principal component analysis}. 
Allen et al. provide a direct way to explore a range of bodies with semantic controls by learning a global linear mapping between the semantic parameters and the PCA weights \cite{Allen2003}.
Yang et al. introduce a local mapping to generate human body models in a variety of shapes and poses via tuning semantic parameters \cite{YangYZDDY14}.
However, these methods rely on global or local feature information to extract the semantic parameters of bodies.
Thus, we present a straightforward and effective method to explore more human bodies without relying on the semantic parameters.
PCA sees the human space as a single, multi-dimensional Gaussian model,
while one can apply more sophisticated analyses (e.g., mixtures of Gaussians) on it to build a more precise statistical model.
Mixtures of Gaussians are often used to segment/co-segment the mesh shapes in computer graphics \cite{Fan2011,cadMengXLH13}. 
We propose a segment-and-merge scheme to implicitly utilize mixtures of Gaussians to model body shapes.
The body mesh is decomposed and recomposed by four semantic components.
Three principal component models including arms, legs and torso are built respectively,
while head is scaled directly according to seam between head and neck.
In Figure \ref{fig:mixture2}, we present an example showing a wide range of leg shapes.
We choose some representive principal components, scale the components according to the ratios between parts and merge them to restore a watertight body again. 
Poisson-based mesh editing approach \cite{Yu2004MEP} is a good candidate for merging.
However, the efficiency is crucial for our task, 
because that tens of thousands of models will be generated by batch processing. 
Straightforwardly,
we scale the regions around the merging seams and smooth them, and then project the merged results into the built shape space.

%%% Figure
%%%

%%% Figure
%%%
\begin{figure}
  \centering
  \hspace{-0.3in}
  \subfigure[]{
    \label{fig:ratio:a} 
    \includegraphics[angle=0,width=2.3in]{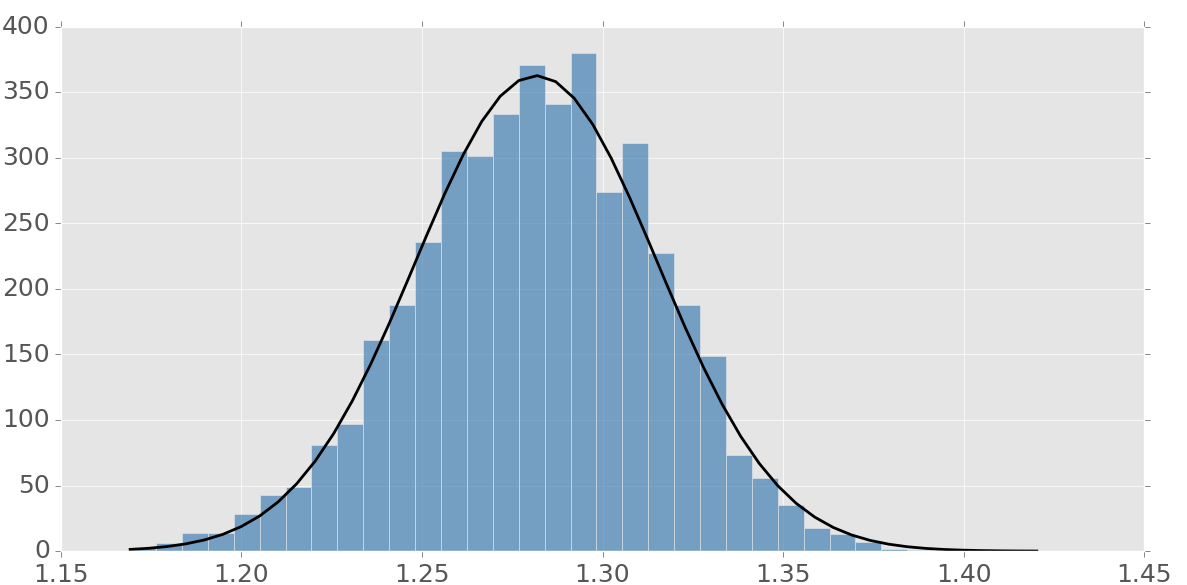}}
  \hspace{-0.1in}
  \subfigure[]{
    \label{fig:ratio:b} 
    \includegraphics[angle=0,width=0.41in]{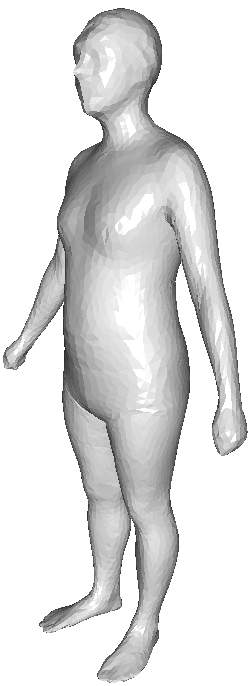}}
    \hspace{-0.10in}
  \subfigure[]{
    \label{fig:ratio:c} 
    \includegraphics[angle=0,width=0.44in]{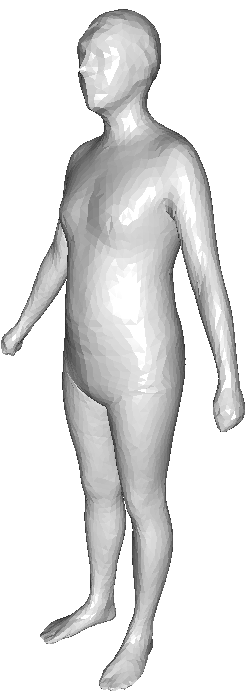}}       
    \hspace{-0.10in}
  \subfigure[]{
    \label{fig:ratio:c} 
    \includegraphics[angle=0,width=0.466in]{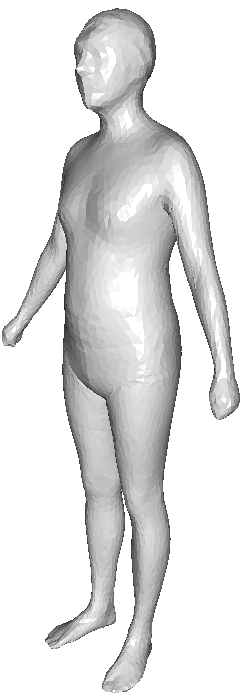}}    
  \caption{\label{fig:ratio}
A Gaussian distribution fits the ratio data quite well. The standard deviation is used to generate new individuals.
  (a) The ratios between legs and torso for 4308 bodies;
  (b-d) three body models generated by merging components with different ratios.
}
\end{figure}

\begin{figure}
  \centering
  \subfigure[]{
    \label{fig:project:a} 
    \includegraphics[angle=0,width=0.744in]{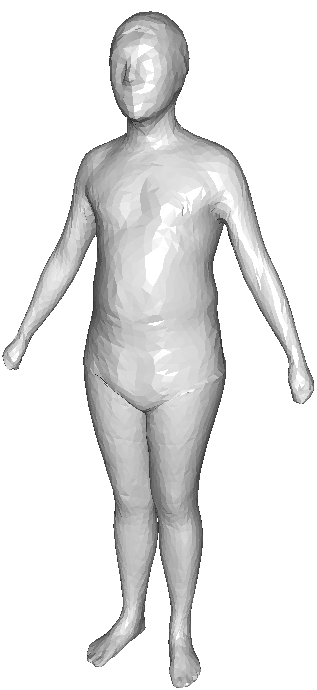}}
  \hspace{-0.05in}
  \subfigure[]{
    \label{fig:project:b} 
    \includegraphics[angle=0,width=0.780in]{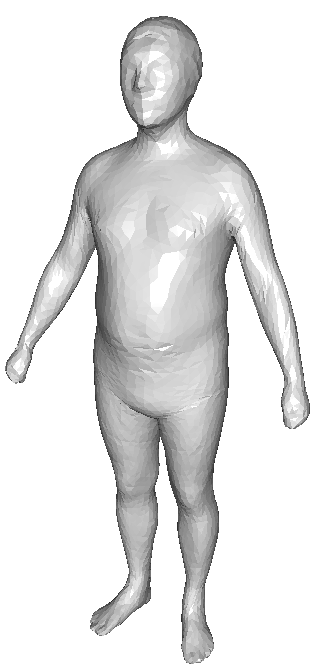}}
    \hspace{-0.05in}
  \subfigure[]{
    \label{fig:project:c} 
    \includegraphics[angle=0,width=0.81in]{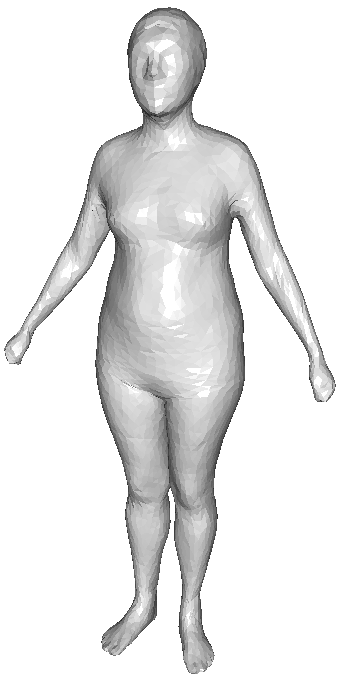}}
    \hspace{-0.05in}
  \subfigure[]{
    \label{fig:project:d} 
    \includegraphics[angle=0,width=0.796in]{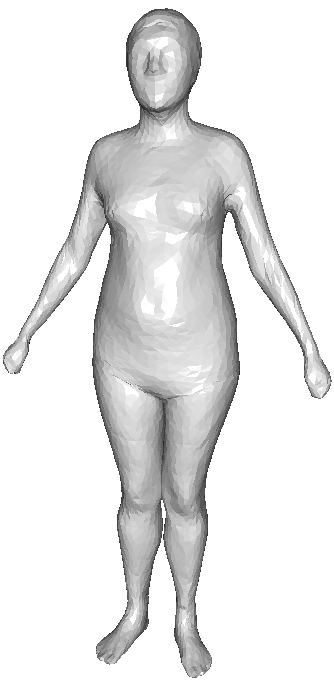}}    
  \caption{\label{fig:project}
An example on projecting into subspaces to improve the shape diversity.
  (a) A body shape created by the segment-and-merge operation;
  (b-d) new individuals generated by projecting (a) into subspaces with $k=\{5,15,30\}$ principal components respectively.
}
\end{figure}

To improve the diversity of body shapes, we leverage the ratios between components.
For example, we measured the ratios between legs and torso for 4308 individuals in the original dataset.
As illustrated in Figure \ref{fig:ratio}, the ratio between components is subject to a Gaussian distribution quite well.
Give a torso model and a leg model, we select 5 ratios $\{\mu-3\sigma,\mu-1.5\sigma,\mu,\mu+1.5\sigma,\mu+3\sigma\}$ to scale leg components, then merge component pairs to generate new individuals.
It is worth noting that, there are two suspended issues about our segment-and-merge scheme.
On the one hand, the simple merging step of our scheme might introduce slight artificials especially near component boundaries,
on the other hand, the resulting body shape might deviate from the built shape space.
We tackle both issues simultaneously by projecting the result back into the shape space built in Eq. (\ref{eqn:pca}). 
This operation plays the role of a regularization which ensures that our scheme produces reasonable results to some extent.
Specially, we project a model into subspaces with $k=\{5,15,30\}$ principal components to obtain some extra individuals at a time. An example is shown in Figure \ref{fig:project}.
Through this technique, more individuals are mined from the shape space built from original 4308 real-world bodies. 

%%% Figure
%%%

Finally, we generate additional 10k models.
In fact, these synthetic bodies span a wide range of global shapes, with the parts stemmed from real-world data. 
It is common knowledge that the more diverse data a deep network has access to, the more effective it becomes.
This strategy for data augmentation further enhances the shape diversity of our body dataset, which guides the deep network learn more variances from the training data.

$\blacksquare$
\textbf{Generative adversarial network}.
In this paper, we will not explicitly build the true distribution of human shape variations,
instead, we intend to exploit the distribution of body shapes using a Generative Adversarial Network (GAN) \cite{NIPS2014_5423}.
GANs are recently developed in the field of deep learning to learn a data distribution and realize a model to sample from it. GANs are architectured around two networks:
the generative network $G(z)$ learns to map from a latent space to the data distribution, while the discriminative network $D(x)$ discriminates between instances from the true data distribution and candidates produced by the generator. The generator and discriminator are typically learned jointly by alternating the training of $D$ and $G$. 
To balance the generator and discriminator during training, and to flexibly control the trade-off between shape diversity and appearance reality, we build a GAN similar to the work \cite{BerthelotSM17}. 
Our generator synthesizes a shape individual $\varphi_s$ from a $30-dimentional$ input sampled from a random uniform distribution and the discriminative network has two competing goals, auto-encode real data and discriminate real from generated data.
We generate extra 15k models to further expand our dataset.

%%% Figure
%%%
\begin{figure}
  \centering
  \hspace{-0.28in}  
  \subfigure[]{
    \label{fig:dataaug:a} 
    \includegraphics[angle=0,width=0.938in]{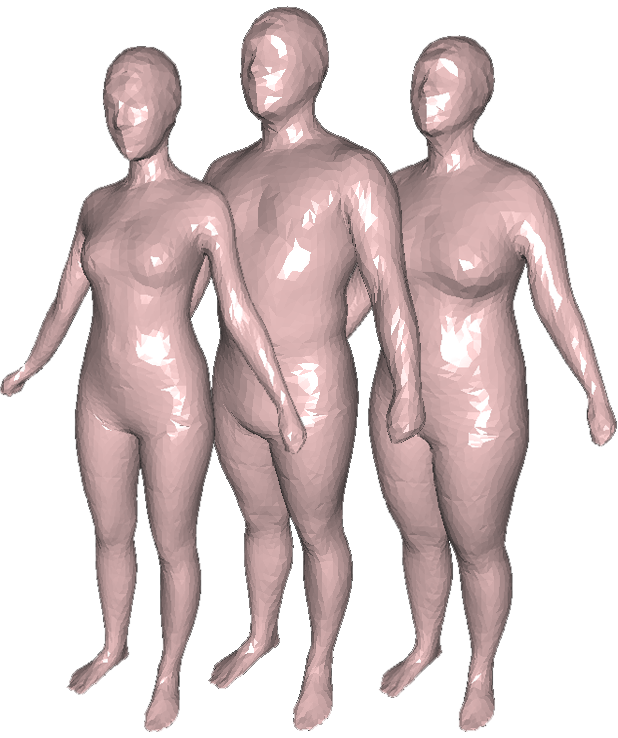}}
  \hspace{-0.15in}
  \subfigure[]{
    \label{fig:dataaug:b} 
    \includegraphics[angle=0,width=0.938in]{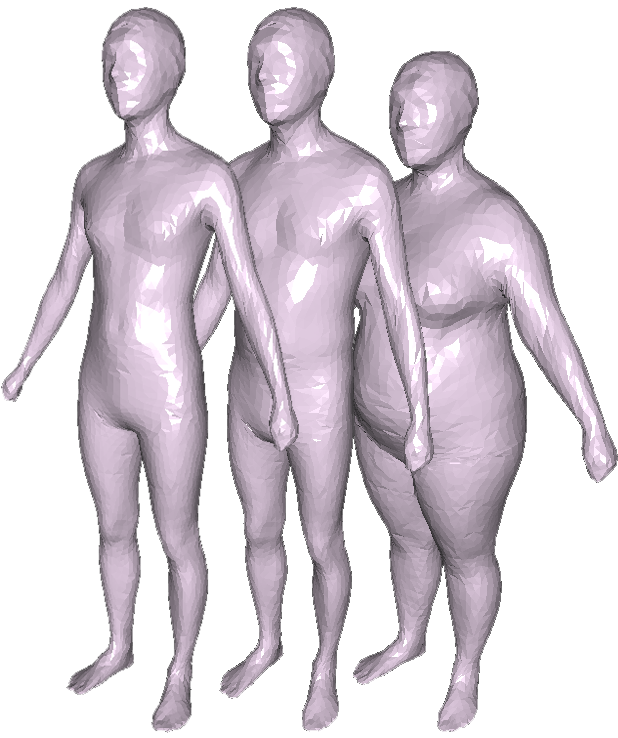}}
    \hspace{-0.15in}
  \subfigure[]{
    \label{fig:dataaug:c} 
    \includegraphics[angle=0,width=0.938in]{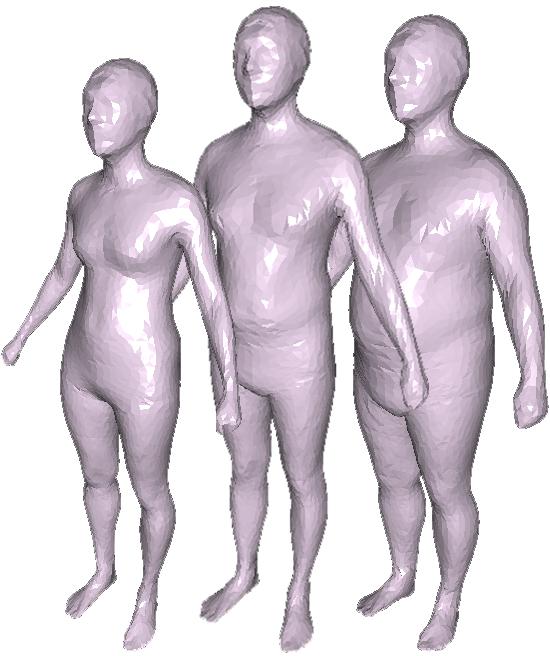}}
    \hspace{-0.15in}
  \subfigure[]{
    \label{fig:dataaug:d} 
    \includegraphics[angle=0,width=0.938in]{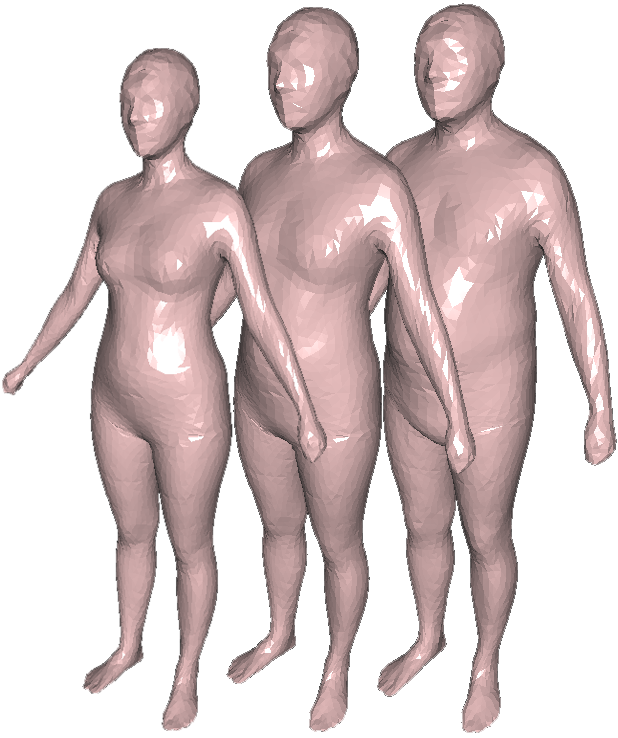}}    
  \caption{\label{fig:dataaug}
A gallery of body models, including some bodies from original dataset and those generated using our strategy for data augmentation.
  (a) Some original body models;
  (b) body models generated by interpolating and random sampling;
  (c) body models generated by merging and projecting into the built shape space;
  (d) body models generated using a GAN.
}
\end{figure}

Some body models are shown in Figure \ref{fig:dataaug}.
The synthetic data have a realistic appearance for the most part, but do not
look the same as any particular individual from the original dataset.
These generated data and the original data together form our final dataset ($\sim$50k models) for network training and testing. 
80\% models of our final dataset are randomly chosen as training data, and 20\% as testing data.

\begin{figure*}
  \centering
  \subfigure{
    \label{fig:network:a} 
    \includegraphics[angle=0,width=7.1in]{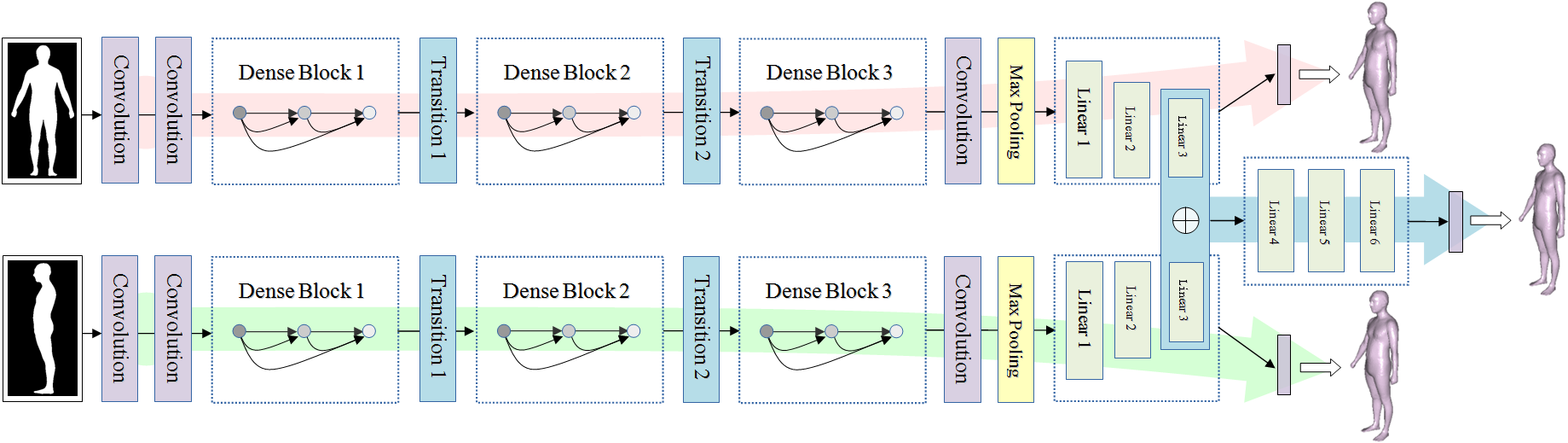}}
  \caption{\label{fig:network}
Illustration of our network architecture for human body reconstruction with two-view masks. Our network is composed of several dense blocks.
It is allowed to perform by passing through two separate pipelines starting from a single frontal mask (along the direction of light red arrow) or from a single lateral mask (along the direction of light green arrow). And the joint pipeline is conducted by concatenating the feature descriptors in the light blue box and goes forward along the direction of light blue arrow.}
\end{figure*}

\section{Human Body Inference from Views}

\subsection{Network architecture}

The architecture of our deep regression network for 3D human body reconstruction from mask images is illustrated in Figure \ref{fig:network}. 
It utilizes multiple dense blockes and transition layers to extract and aggregate feature maps from input masks.

The main motivation behind our architecture is that our 3D model inference is performed by gathering local and global structure information. Specifically speaking,
the network should have an opportunity to process features at both
local and global contexts without upsampling layers, and neurons in the last convolutional layer possess a large receptive fields to grasp the global contextual information of an input mask. 
To this end, we build a network with several dense blocks which connects each layer to every other layer in a feed-forward fashion.
Dense block is introduced by DenseNet which has shown excellent results on computer vision tasks, such as image classification and object recognition. 
For each layer in a dense block, the feature-maps of all preceding layers are used as inputs, and its own feature-maps are inputed into all subsequent layers.
This fashion of architecture facilitates feature propagation, encourages feature reuse, and substantially reduces the number of model parameters. 

As illustrated in Figure \ref{fig:network},
we design such a novel CNN with two branches to decouple the features of two-view masks 
by training a separate network for each mask respectively.
In our architecture, two branches in the first part of the network do not share the model parameters and even may possess different structures.
Such a design is to represent features of two-view masks using different descriptors respectively.
Considering the flexibility of the algorithm, 
our model predictions are allowed to perform by passing through three pipelines, 
including two separate pipelines starting from a frontal mask or a lateral mask, and the joint pipeline.
Due to the correlation between two views of a body shape, 
given a single mask in a standard posture, our approach is still able to estimate a plausible body parameters.
It becomes useful if only a single mask is given.
Through the joint pipeline, it will generate more accurate result if two masks captured from the same body are provided.
Additionally, the user can easily obtain some variants by editing one or two masks slightly.
We call these three networks as FrontalNet, LateralNet and JointMaskNet respectively.

\subsection{Reconstruction with two-view masks}
It is commonly known that 3D modeling via a 2D image plane is a time-cosuming task.
Methods through illustrating a set of 2D images from orthogonal views do really benefit it.
Before modeling new geometries, 3D artists tend to exhibit their concept design through assigning three-view 2D images. 
Inspired by this modeling style, this paper focuses on 3D modeling of human body with one or two images in orthogonal views. 
Besides the silhouettes, mask image encodes local and global shapes captured from different views.
Mask images can be estimated from color images or constructed from scratch directly with the interactive user aid indicated via sketching or brushing.
Convolutional neural networks are trained to estimate and recover the 3D body shape from these shape cues.

For human body modeling, a top-view conveys limited cues because of the self-occlusion of body parts.
Fortunately, front-view and side-view can adequately convey the main shape of a human body.
This paper proposes a novel method for rapid reconstruction of 3D human body model from two-view masks.
The proposed method is able to learn the complementary information from two views using a deep network architecture with two branches. 
In our architecture, there are two independent deep networks, and all the parameters of each deep network are learned discriminatively to produce a single-view descriptor for the 3D shape. 
Two single-view descriptors can be concatenated together to estimate the final 3D body shape.
As mentioned above, two branches in our architecture does not share the parameters, 
which makes the braches learn their own representive features.

\section{Implementation details}
Our shape-from-mask modeling pipeline starts from mask images of a 3D body shape in orthogonal views.
For each shape in the dataset, we capture and save two binary mask images using the render-to-texture technique in OpenGL.
When rendering the lateral mask, we remove the arms and right leg from the model to obtain a clean lateral profile.
Rendering each 3D model from two viewpoints is very efficient on modern graphics hardware.
For our task, inputed images with a higher resolution provide limited improvments because that fine scale details are absent in a binary mask image.
And the involved convolution network may suffer high computation complexity during network training.
In our experiment, we find $128\times128$ binary mask images perform well.

\textbf{Architecture details} 
In our experiments, first two convolution layers have $11\times11$ and $7\times7$ convolution kernels repectively with a stride of $1$. 
They convert the pixel intensities in mask image to local feature detectors that are then used as inputs to the immediate dense blocks.
A Dense Block is defined as the repetition of a block composed of Batch Normalize, followed by ReLU and convolution layers. 
For convolutional layers in each dense block, the kernel size is $5\times5$, with a stride of $1$ and a padding size of $2$.
Thus the convolutional layers in the same dense block keep a fixed size.
There are two branches in our network, with $3$ dense blocks each of them with $3$ layers and a growth rate of feature maps $12$.
A Transition layer is introdued to reduce the spatical dimensionality of the feature maps.
This transformation layer is composed of Batch Normalize, ReLU and a covolution layer with kernel size of $1\times1$ followed by a $2\times2$ Maxpooling layer.
At the end of each independent branch, there are three fully connected layers before outputing the parameters.
The last fully connected layer of each branch has $512$ neurons. 
And we concatenate the two feature vectors together to form a $1024$ dimentional feature descriptor that is input to another three
fully connecte layers to estimate the terminal parameters.
All layers use ReLU activations except the final layer.

Our 3D shape-from-mask reconstruction method is render-free, without optimization loops which render the reconstructed shape and compare it to the input mask,
and the loss function is defined as follows,
\[ 
L(\varphi,\varphi_s) =  (1-\lambda)||\varphi - \varphi_s||^2 + \lambda ||\Omega (\varphi - \varphi_s)||^2 / N
\]
where $N$ is the total number of vertices.
To make the training procedure converge stably, we take a simple but effective training strategy
by setting $\lambda=0$ initially and then increasing it gradually.
Each branch of our model is trained separately with the parameters randomly initialized.
When both branches are learned, we fix them and train the remaining network, JointMaskNet.
Specifically, we extract the outputs of two branches for every training data 
and concatenate them together offline, as a composit feature descriptor, 
to train the joint network comprised of the last three fully connected layers until convergence.
We implement our network in Pytorch and train it on a single Titan 1080ti GPU.

\section{Experimental results}
Given that this work focus on body shape reconstruction with a standard posture, utilizing the deep learning technique.
We intend to evaluate the reconstruction accurary of our approach on a subset of the full testing set (about 1000 real world human bodies randomly chosen from the CAESAR dataset).
Now we evaluate and compare our proposed networks, including FrontalNet, LateralNet and JointMaskNet.

\begin{figure}
  \centering
  \hspace{-0.128in}  
  \subfigure[]{
    \label{fig:singleview0:a} 
    \includegraphics[angle=0,width=0.7399in]{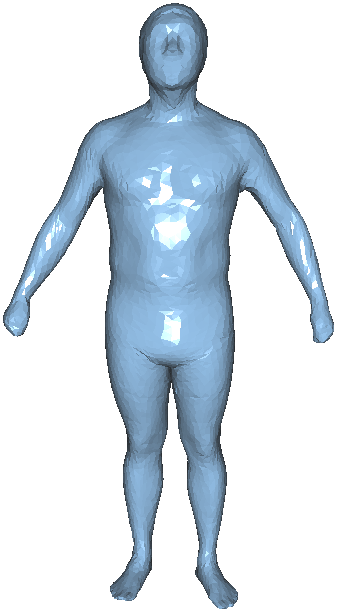}}
  \hspace{-0.09in}
  \subfigure[]{
    \label{fig:singleview0:b} 
    \includegraphics[angle=0,width=0.7399in]{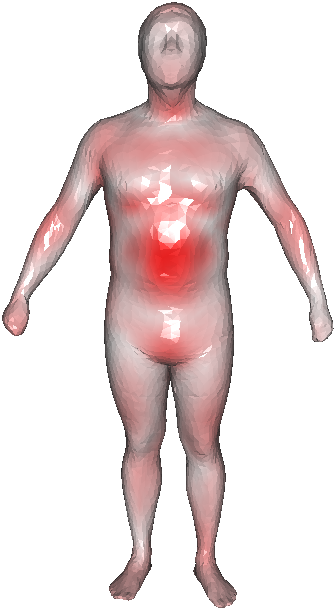}}
    \hspace{-0.09in}
  \subfigure[]{
    \label{fig:singleview0:c} 
    \includegraphics[angle=0,width=0.7399in]{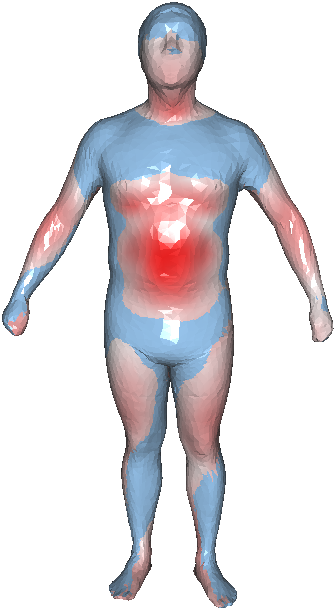}}
    \hspace{-0.09in}
  \subfigure[]{
    \label{fig:singleview0:d} 
    \includegraphics[angle=0,width=0.6in]{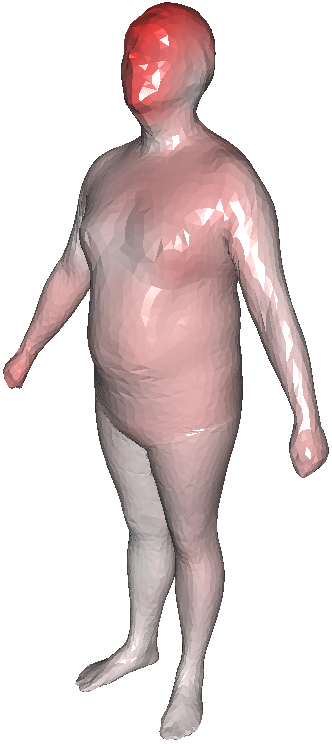}}
    \hspace{-0.09in}
  \subfigure[]{
    \label{fig:singleview0:e} 
    \includegraphics[angle=0,width=0.6in]{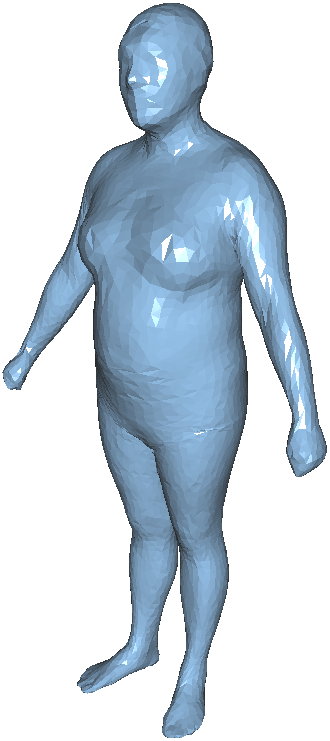}}        
  \caption{\label{fig:singleview0}
Illustration of two examples generated by FrontalNet. 
  (a) Ground truth;
  (b) result of FrontalNet, rendering with an error map (redder indicates higher error);
  (c) overlap of (a) and (b); (d) another result generated by FrontalNet; (e) ground truth of (d).
}
\end{figure}

\textbf{Visual results of 3D human shapes with a single mask.} 
We begin by demonstrating the visual results of FrontalNet.
Figure \ref{fig:singleview0} gives two examples to illustrate the effectiveness of FrontalNet.
In this figure, the bodies are generated by a single frontal mask image. 
At a glance, our FrontalNet produces plausible results which capture the global shape and the proportions between parts of human body. 
From a frontal mask, the FrontalNet can recover the variations along the orientation perpendicular to the camera.
However, in most cases, a single frontal mask is not sufficient to capture an entirely accurate body shape. 
Specifically, incorrect estimates often occur in some areas, 
such as the head, the breasts and the abdomen in a frontal image. 
Because the frontal mask is hard to encode the variations of a body along the orientation towards the camera.
Thus a single frontal mask is incapable of constraining the body shape in these regions,
which induces the ambiguity.
As can be seen in Figure \ref{fig:singleview0}, the abdomen in Figure \ref{fig:singleview0:b} and
the head in Figure \ref{fig:singleview0:d} exhibit some highest errors.
Extra constraints should be added to decrease the degree of freedom in these regions.

\begin{figure}
  \centering
  \hspace{-0.128in}  
  \subfigure[]{
    \label{fig:singleview1:a} 
    \includegraphics[angle=0,width=0.688in]{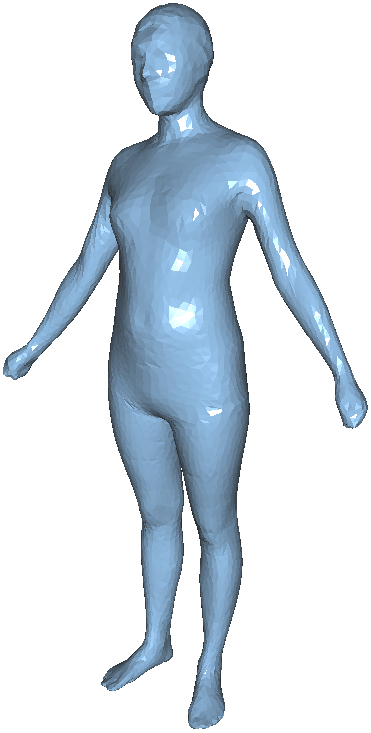}}
  \hspace{-0.1in}
  \subfigure[]{
    \label{fig:singleview1:b} 
    \includegraphics[angle=0,width=0.666in]{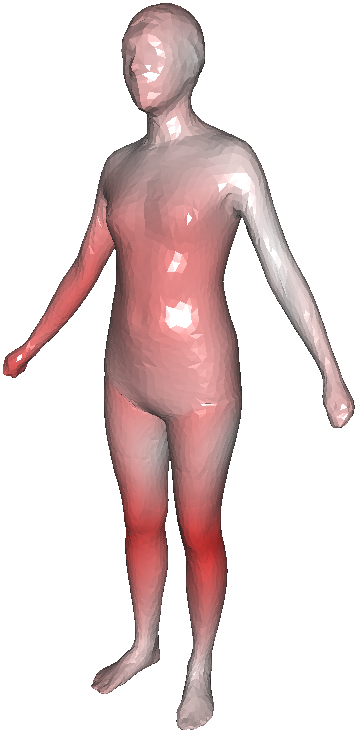}}
    \hspace{-0.1in}
  \subfigure[]{
    \label{fig:singleview1:c} 
    \includegraphics[angle=0,width=0.688in]{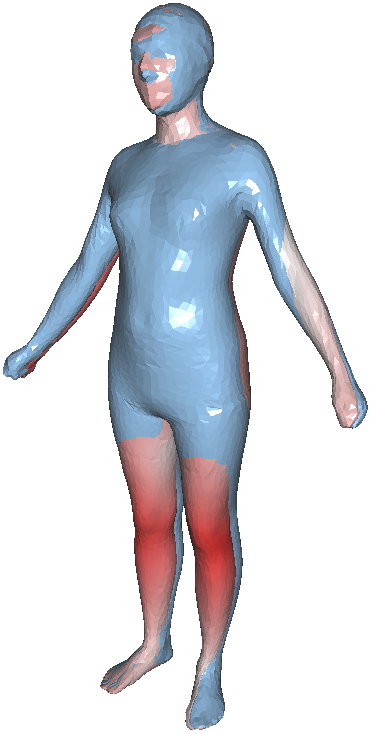}}
    \hspace{-0.1in}
  \subfigure[]{
    \label{fig:singleview1:d} 
    \includegraphics[angle=0,width=0.662in]{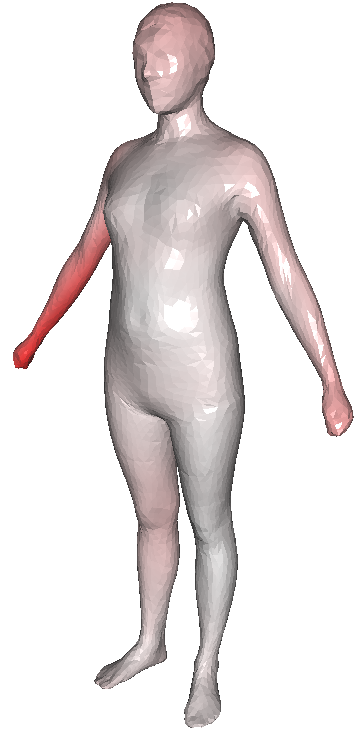}}
    \hspace{-0.1in}
  \subfigure[]{
    \label{fig:singleview1:e} 
    \includegraphics[angle=0,width=0.69in]{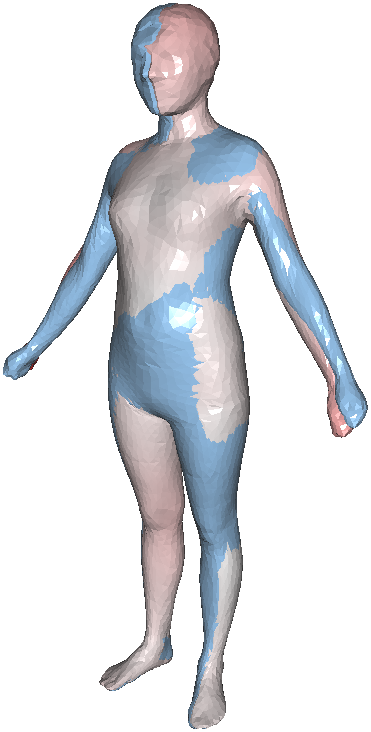}}        
  \caption{\label{fig:singleview1}
An example produced using a single mask image and compared with the ground truth. 
  (a) Ground truth; (b) result of FrontalNet; (c) overlap of (a) and (b); (d) result of LateralNet; (e) overlap of (a) and (d).
}
\end{figure}

In Figure \ref{fig:singleview1}, we provide an additional example that illustrates the effectiveness 
and limitation of reconstruction from a single mask image.
It shows the errors of FrontalNet's and LateralNet's predictions against the ground truth.
For example, FrontalNet produces a body with upright legs, while the legs of ground truth bend backwards slightly. 
Observing from the left side of the body, LateralNet predicts the shapes of legs more accurately, whereas the arms and the posture are its blind spots.
As can be seen in Figure \ref{fig:singleview0} and Figure \ref{fig:singleview1}, 
a single mask can capture the global shape and body proportions to some extent,
but it may introduce significant errors in some regions.
In these figures, the reconstructed human bodies are rendered with an error map indicating per vertex displacement from ground truth, 
white indicates low error while red indicates high error.

%%% Figure
%%%

%%% Figure
%%%

\begin{figure}
  \centering
  \hspace{-0.158in}  
  \subfigure[]{
    \label{fig:doubleview0:a} 
    \includegraphics[angle=0,width=0.705in]{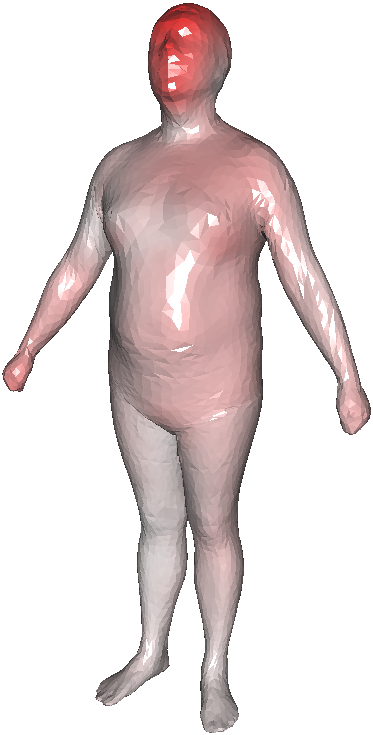}}
    \hspace{-0.1in}
  \subfigure[]{
    \label{fig:doubleview0:b} 
    \includegraphics[angle=0,width=0.7in]{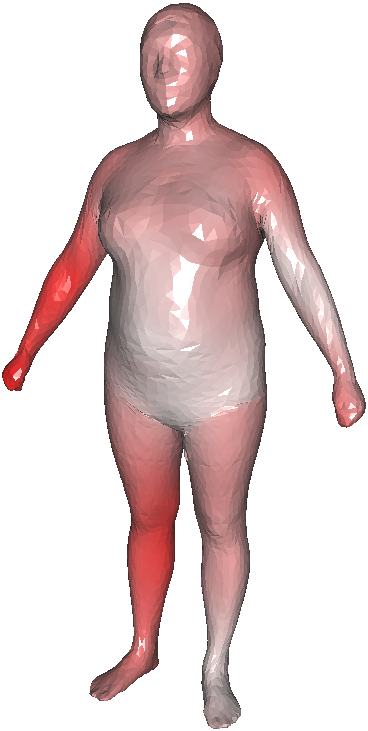}}
    \hspace{-0.1in}
  \subfigure[]{
    \label{fig:doubleview0:c} 
    \includegraphics[angle=0,width=0.695in]{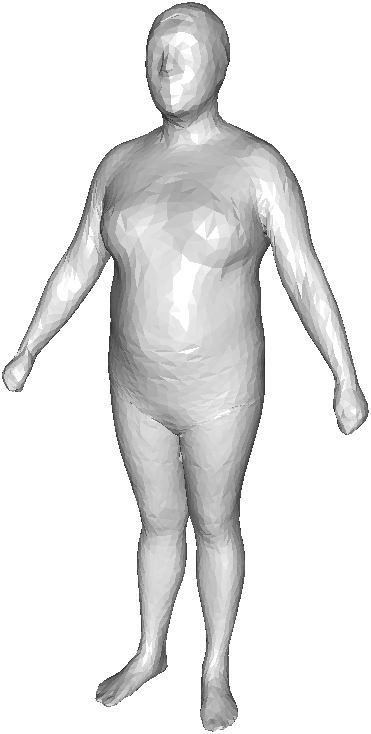}}
    \hspace{-0.1in}
  \subfigure[]{
    \label{fig:doubleview0:d} 
    \includegraphics[angle=0,width=0.7in]{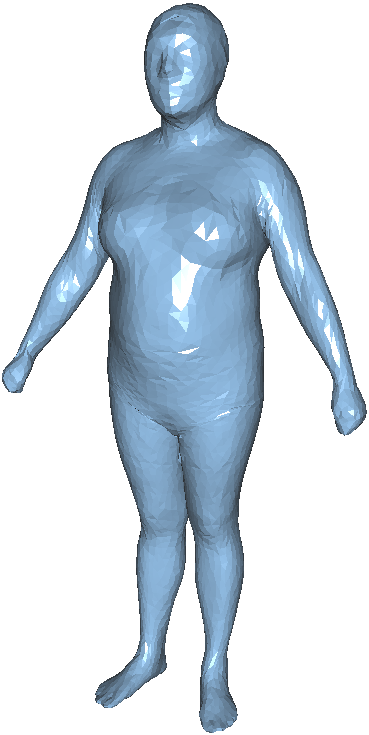}}
    \hspace{-0.1in}
  \subfigure[]{
    \label{fig:doubleview0:e} 
    \includegraphics[angle=0,width=0.7in]{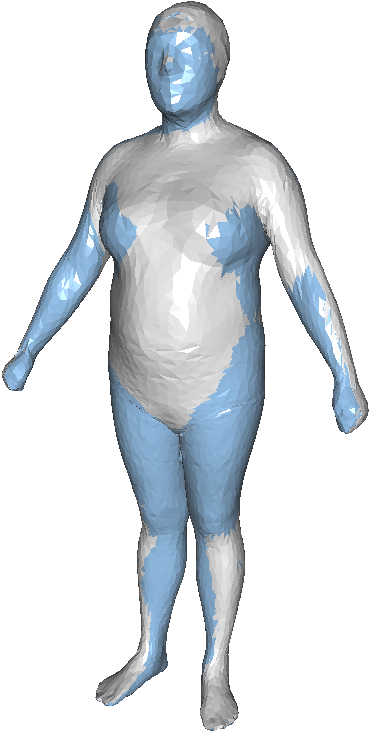}}        
  \caption{\label{fig:doubleview0}
Comparison of the results of three networks and the ground truth.
  (a) result of FrontalNet; (b) result of LateralNet; (c) result of JointMaskNet; (d) ground truth; (e) overlap of (c) and (d).
}
\end{figure}

\begin{figure}
  \centering
  \hspace{-0.18in}  
  \subfigure[]{
    \label{fig:doubleview1:a} 
    \includegraphics[angle=0,width=0.72in]{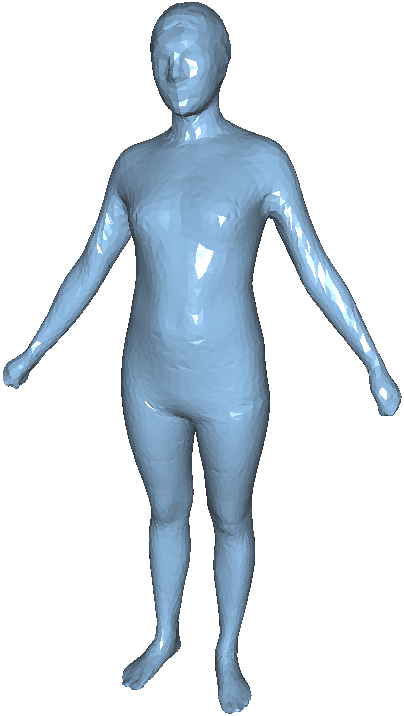}}
    \hspace{-0.08in}
  \subfigure[]{
    \label{fig:doubleview1:b} 
    \includegraphics[angle=0,width=0.72in]{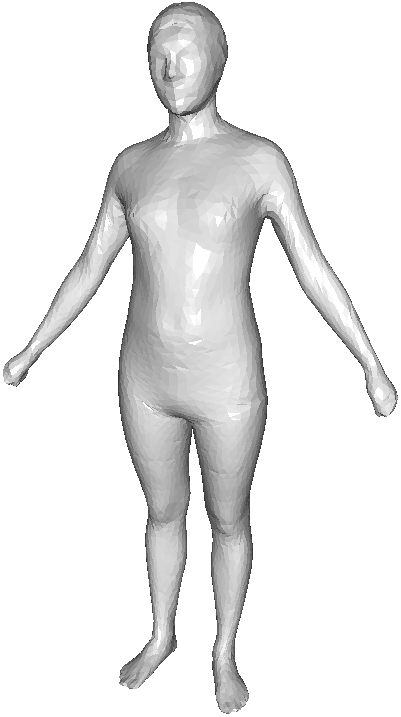}}
    \hspace{-0.08in}
  \subfigure[]{
    \label{fig:doubleview1:c} 
    \includegraphics[angle=0,width=0.72in]{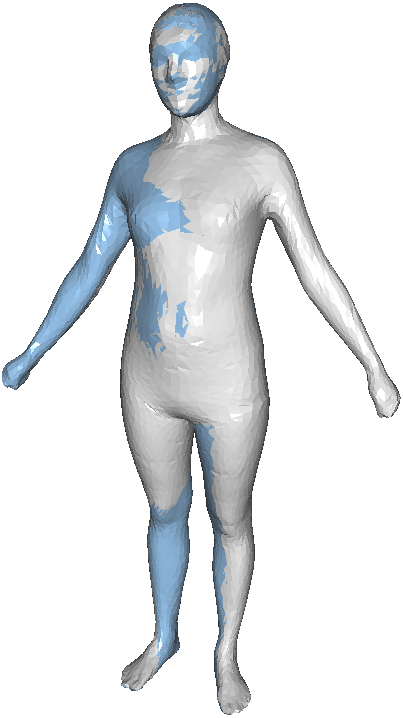}}
    \hspace{-0.08in}
  \subfigure[]{
    \label{fig:doubleview1:d} 
    \includegraphics[angle=0,width=0.655in]{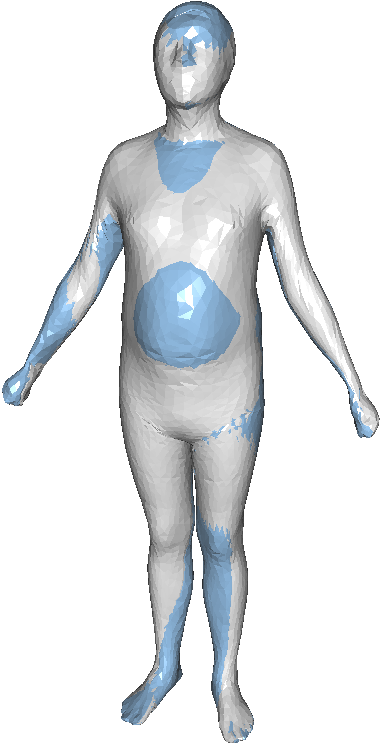}}
    \hspace{-0.08in}
  \subfigure[]{
    \label{fig:doubleview1:e} 
    \includegraphics[angle=0,width=0.615in]{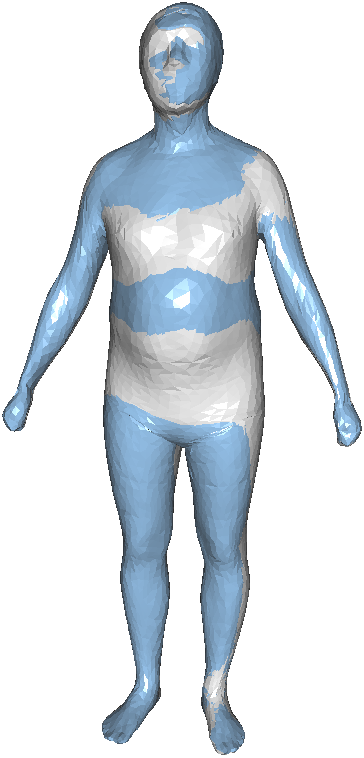}}        
  \caption{\label{fig:doubleview1}
Some results generated by our joint network. 
  (a) Ground truth; (b) our result; (c) overlap of (a) and (b); (d,e) more examples.
}
\end{figure}

\textbf{Visual results of 3D human shapes with two-view masks.} 
To improve the quality of reconstruction, two independent but complementary strategies should be coupled together.
Accordingly, we intend to provide a straightforward and effective solution to take advantage of both views. 
We first train two networks separately to decouple the features from two masks,
resulting in two separate descriptors which encode the body parameters from a certain view respectively.
And then we concatenate the extracted feature vectors to reconstruct an entire body shape.
Figure \ref{fig:doubleview0} provides a visual comparison of the shape errors for three networks.
Figure \ref{fig:doubleview0:a} shows that FrontalNet captures most body parts but the head posture and the breast which are captured well by LateralNet (see Figure \ref{fig:doubleview0:b}).
However, LateralNet fails to accurately recover the postures of the arm and leg on the right side, 
which is due to the fact that they are invisible from the left side.
The above facts suggest that the two views are complementary to some extent.
Figure \ref{fig:doubleview0:c} presents the result of our JointMaskNet, showing more approximate to
the groud truth (see Figure \ref{fig:doubleview0:d}).
In Figure \ref{fig:doubleview0:e}, we put them together to display the indistinguishable errors on overlapped regions more clearly.
Most regions are recovered wery well and induce the z-fighting issue in OpenGL,
but in some areas, such as the abdomen and thighs, there exist some tiny displacements.
More examples are shown in Figure \ref{fig:doubleview1} to illustrate the improvement of the reconstructed accuracy from JointMaskNet. 
Our JointMaskNet produce more accurate results due to the removal of most ambiguities existing in a single-view mask.
Note that these body shapes are randomly chosen from the testing set.

%%% Figure
%%%

%%% Figure
%%%

\begin{figure}
  \centering
  \hspace{-0.05in}
  \subfigure[]{
    \label{fig:error:a} 
    \includegraphics[angle=0,width=1.13in]{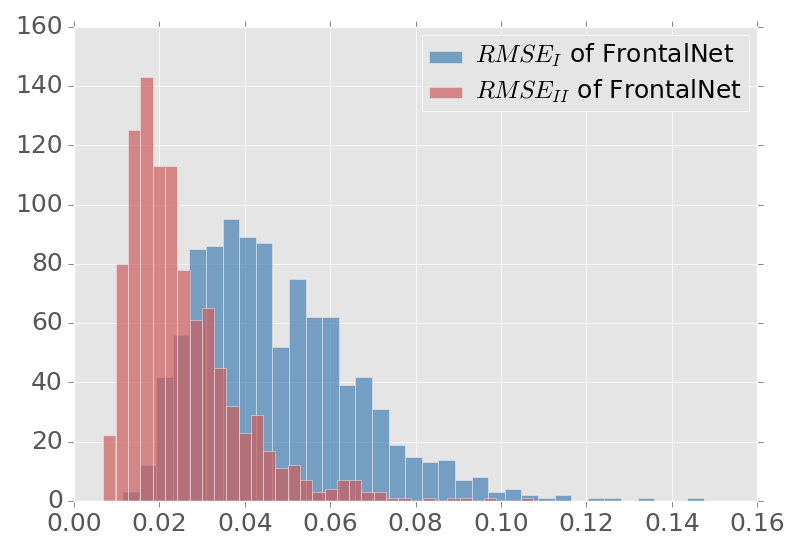}}
  \hspace{-0.105in}
  \subfigure[]{
    \label{fig:error:b} 
    \includegraphics[angle=0,width=1.13in]{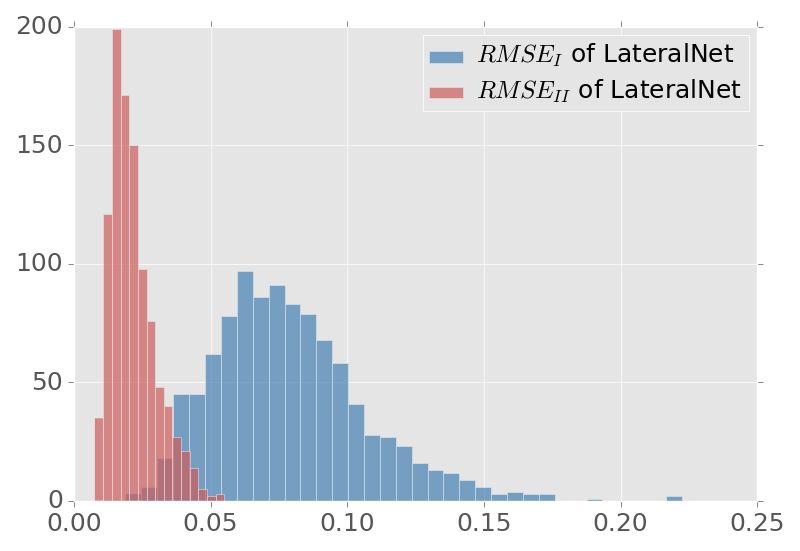}}
    \hspace{-0.105in}
  \subfigure[]{
    \label{fig:error:c} 
    \includegraphics[angle=0,width=1.13in]{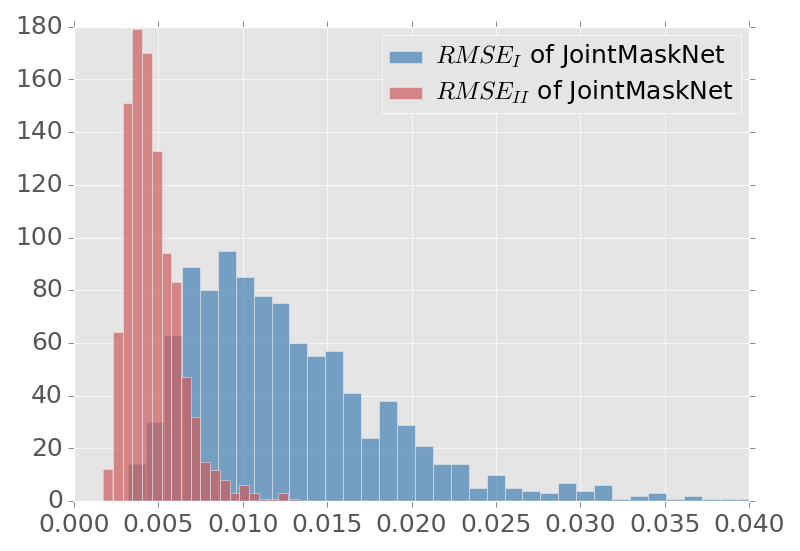}}
  \hspace{-0.0in}    
  \subfigure[]{
    \label{fig:error:d} 
    \includegraphics[angle=0,width=1.12in]{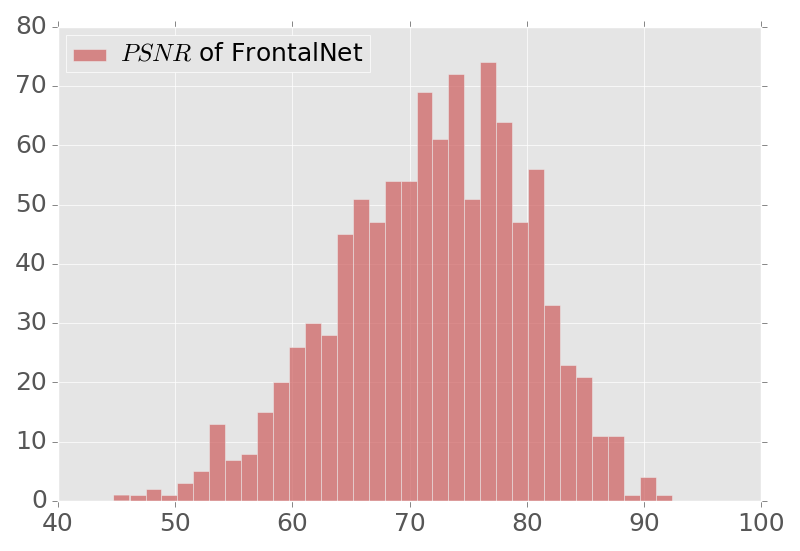}}
  \hspace{-0.105in}
  \subfigure[]{
    \label{fig:error:e} 
    \includegraphics[angle=0,width=1.12in]{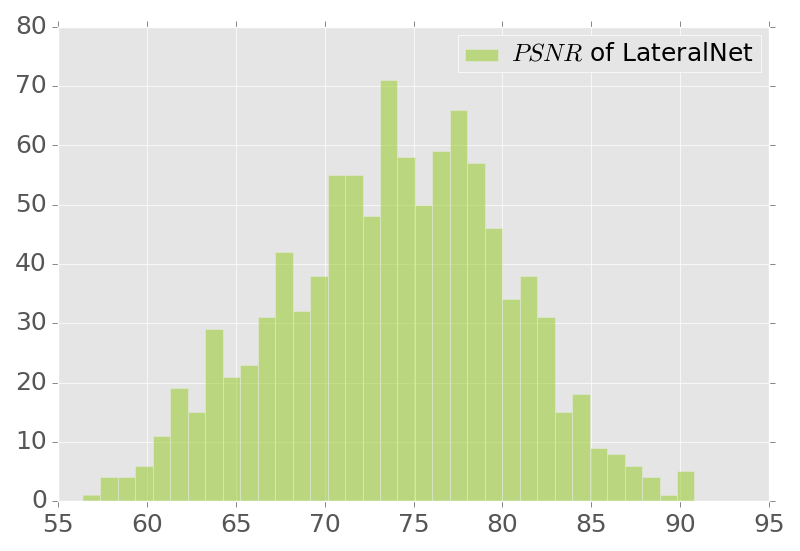}}
    \hspace{-0.105in}
  \subfigure[]{
    \label{fig:error:f} 
    \includegraphics[angle=0,width=1.12in]{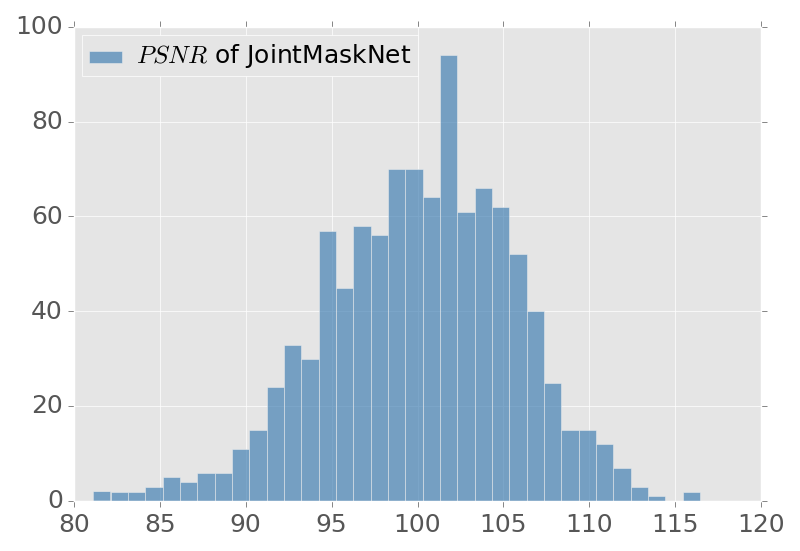}}    
  \caption{\label{fig:error}
Qualitative illustration of geometric errors and PSNRs for three networks. The error distributions in real world meters $m$ are shown. We assume that the average height of adult human bodies is 1.70$m$, and we demonstrate the distributions of the point-to-point maximum error/mean error in the testing set (about 1000 real-world bodies). 
  (a) FrontalNet; (b) LateralNet; (c) JointMaskNet; (d) PSNR of FrontalNet; (e) PSNR of LateralNet; (f) PSNR of JointMaskNet. 
}
\end{figure}

\textbf{Qualitative evaluation of 3D shape reconstruction accuracy.}
We now perform a qualitative evaluation the accuracy of its reconstructed 3D shape on some existing real-world 3D body shapes with standard posture.
We randomly select some human bodies from CAESAR dataset to evaluate our reconstruction accurary.
This testing set contains about 1000 bodies which spans a wide range of individuals.
For visualizing the errors, we use three metrics, Root Maximum Square Error ($RMSE_{I}$), 
Root Mean Square Error ($RMSE_{II}$) and Peak Signal-to-Noise Ratio ($PSNR$),
to measure the difference between the reconstructed shape and the ground truth in the term of vertex.
The root maximum square error measuring the largest displacement in the term of vertex, is defined by $max_{i}\{||v_i-\hat{v}_i||_2\}$ where $v_i$ and $\hat{v}_i$ are the $i$-th vertex of the ground truth and a reconstructed body shape respectively.
The root mean square error representing the cumulative squared error between the reconstructed and the original model, is evaluated by computing $RMSE_{II} = \sum_{i=1}^{N}\{||v_i-\hat{v}_i||_2\}/N$.
We also compute the $PSNR$ to measure the fidelity of reconstruction, using the equation $PSNR = 20log_{10}(MAX_Z)-10log_{10}(MSE)$, where $MAX_Z$ is the range of $Z$ values of body model, and $MSE$ is the square of $RMSE_{II}$.
Figure \ref{fig:error} provides a qualitative accuracy comparison of reconstruction errors in real world meters $m$ for three networks on the testing dataset. 
The JointMaskNet introduces smallest errors among three networks compared to ground truth. 
For FrontalNet, the posture of head and legs, the regions of the breast and abdomen may introduce more errors than other regions.
Obviously, the LateralNet generates the largest maximum errors which are introduced by the arms and legs in particular because of self-occlusions.
$PSNR$ and $RMSE_{II}$ evaluate the global errors over the entire body shape, 
whereas $RMSE_{I}$ only captures the largest error which is usually introduced by the posture.
Figure \ref{fig:error} demonstrate that $RMSE_{II}$s of FrontalNet and LateralNet are much lower than their $RMSE_{I}$s, which is acceptable for some applications if only a single mask is available.
Remarkablely, the JointMaskNet is more accurate than the single mode, 
with lower $RMSE_{I}$, $RMSE_{II}$ values and higher PSNR values.
This may be due to the complementarity between two masks and our use of a large dataset to train a CNN-based network.
FrontalNet focuses on capturing the posture of limbs and the variations in the coronal plane, 
and LateralNet estimates the posture of head and the variations in the sagittal plane, 
while JointMaskNet couples them together to recover the body shape in a more comprehensive manner.

%%% Figure
%%%

\begin{figure}
  \centering
  \hspace{-0.158in}  
  \subfigure[]{
    \label{fig:composition0:a} 
    \includegraphics[angle=0,width=1.05in]{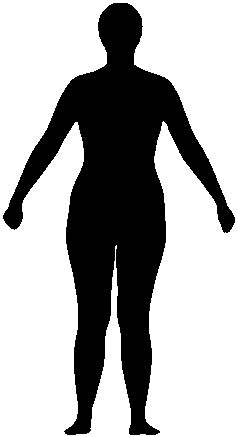}}
    \hspace{0.015in}
  \subfigure[]{
    \label{fig:composition0:b} 
    \includegraphics[angle=0,width=0.344in]{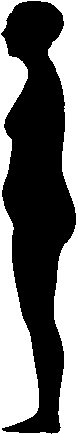}}
    \hspace{0.015in}
  \subfigure[]{
    \label{fig:composition0:c} 
    \includegraphics[angle=0,width=0.344in]{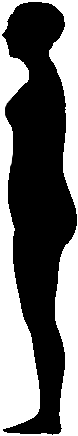}}    
    \hspace{0.1in}
  \subfigure[]{
    \label{fig:composition0:d} 
    \includegraphics[angle=0,width=0.338in]{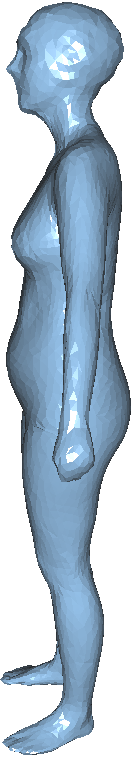}}
    \hspace{0.05in}
  \subfigure[]{
    \label{fig:composition0:e} 
    \includegraphics[angle=0,width=0.34in]{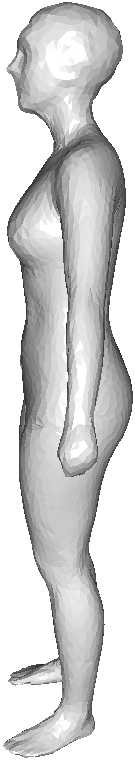}}        
    \hspace{0.05in}
  \subfigure[]{
    \label{fig:composition0:f} 
    \includegraphics[angle=0,width=0.341in]{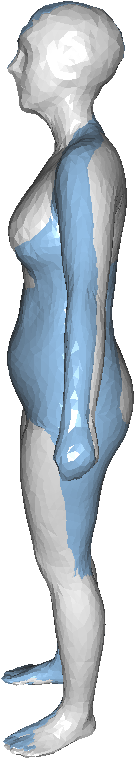}}      
  \caption{\label{fig:composition0}
Modifications of a lateral mask produce bodies with different local shapes. 
We intend to boost the breast and to reduce the lower abdomen by editing a lateral mask manually.
The modification of mask is just intuitive and inaccurate, and the resulting body fits the edited profile roughly.
  (a) Frontal mask; (b) lateral mask; (c) the edited lateral mask; (d) the original body;
  (e) the resulting body from (a) and (c); (f) the overlap of (d) and (e).
}
\end{figure}

\textbf{Generalization of JointMaskNet.}
We have demonstrated the generalization ability of our network on a testing set composed of real world data,
we next perform an interesting experiment on our JointMaskNet to estimate its generalization for inconsistent mask images.
We aim to see if our network works according to our modification imposed on one or two mask images.
The masks may be edited manually or be captured from different persons.
Figure \ref{fig:composition0} gives an example to show the effect when we modify one mask.
For example, we want to change the shapes of breast and lower abdomen of an existing body.
Given the lateral mask image, we first edit the target regions on the image intuitively.
Then our JointMaskNet produces a new body shape according to two masks.  
The resulting body might not fit the edited profile accurately.
The reason for this is twofold: first, the edited masks may be unreasonable, 
and second, our JointMaskNet is restricted in the body shape space which plays a role of regularization. 
Anyhow, our method can be used to generate a series of visually plausible 3D body shapes 
which are similar to a existing body shape.

\begin{figure}
  \centering
  \hspace{-0.158in}  
  \subfigure[]{
    \label{fig:composition1:a} 
    \includegraphics[angle=0,width=0.85in]{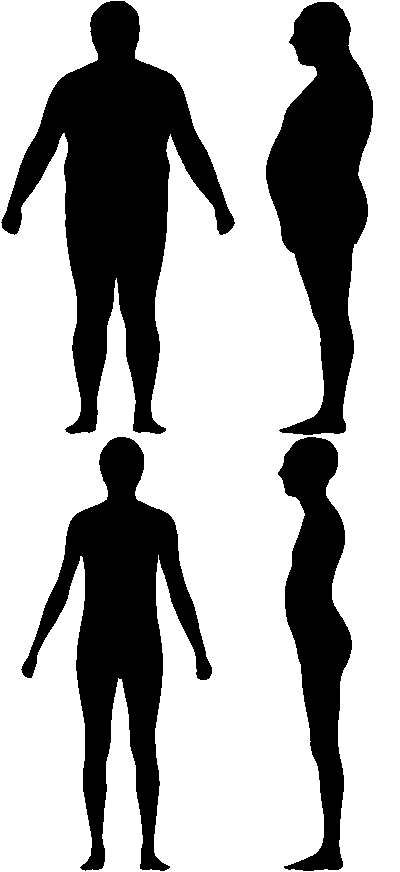}}
    \hspace{-0.05in}
  \subfigure[]{
    \label{fig:composition1:b} 
    \includegraphics[angle=0,width=0.925in]{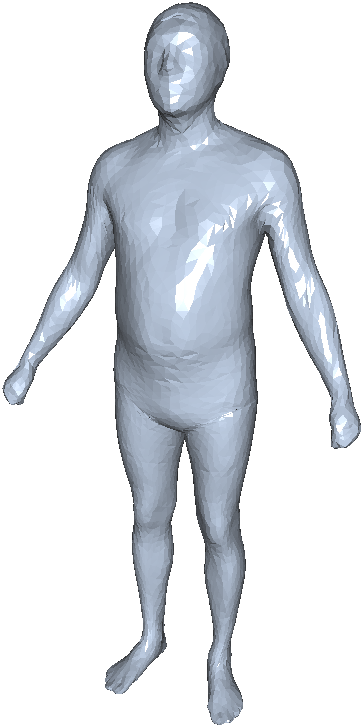}}
    \hspace{-0.1in}
  \subfigure[]{
    \label{fig:composition1:c} 
    \includegraphics[angle=0,width=0.366in]{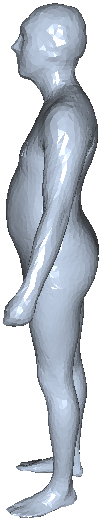}}
    \hspace{-0.0in}
  \subfigure[]{
    \label{fig:composition1:d} 
    \includegraphics[angle=0,width=0.793in]{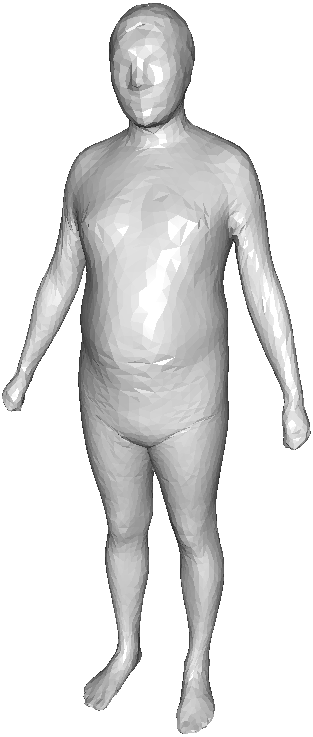}}        
    \hspace{-0.1in}
  \subfigure[]{
    \label{fig:composition1:e} 
    \includegraphics[angle=0,width=0.4in]{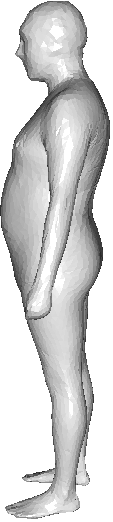}}      
  \caption{\label{fig:composition1}
Illustration of an example generated by exchanging the mask images of two subjects. 
The involved mask images are packaged into a single image.
  (a) Masks of subject A (upper row) and B (lower row); (b,c) the first result; (d,e) the second result.
}
\end{figure}

What happens when our JointMaskNet takes the input of two masks sampled from different persons?
Figure \ref{fig:composition1} presents such an example.
In particular, we exchange the masks of two persons with remarkably different shapes (a fatter person A and a thinner one B),
and obtain two body shapes by applying JointMaskNet twice. 
The first resulting body shape exhibits a frontal profile similar to subject A, and a lateral profile similar to subject B,
whereas the second resulting body is exactly reversed.
When taking the input of inconsistent masks, JointMaskNet might be confused and produces a compromised result to match both profiles as much as possible.
Because its output is restricted in the built body space, 
which makes the resulting shape be a normal body even given two inconsistent masks.

%%% Figure
%%%

%%% Figure
%%%

\begin{figure}
  \centering
  \hspace{-0.258in}  
  \subfigure[]{
    \label{fig:hd0:a} 
    \includegraphics[angle=0,width=0.795in]{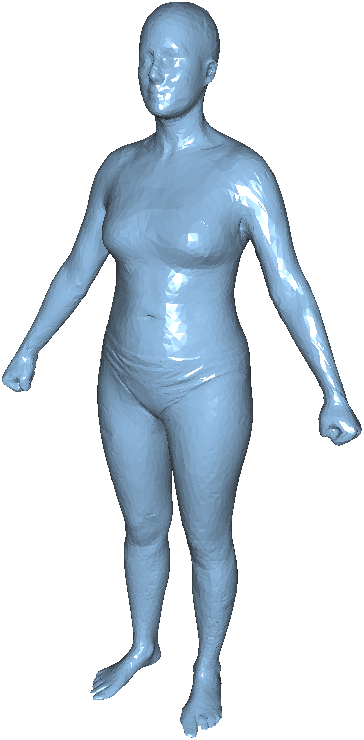}}
    \hspace{-0.115in}
  \subfigure[]{
    \label{fig:hd0:b} 
    \includegraphics[angle=0,width=0.43in]{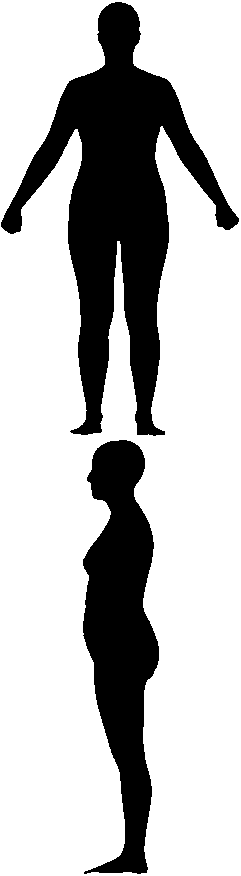}}
    \hspace{-0.115in}
  \subfigure[]{
    \label{fig:hd0:c} 
    \includegraphics[angle=0,width=0.78in]{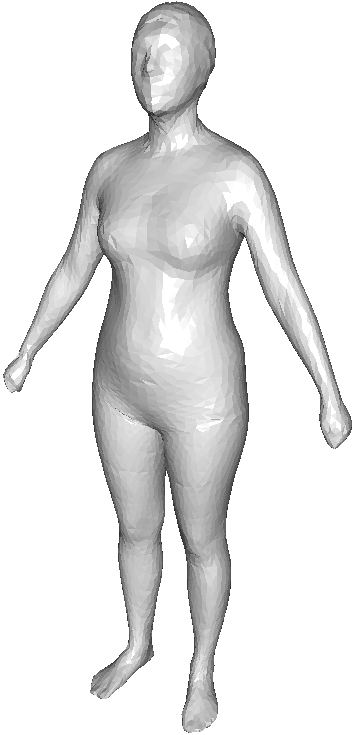}}
    \hspace{-0.115in}
  \subfigure[]{
    \label{fig:hd0:d} 
    \includegraphics[angle=0,width=0.78in]{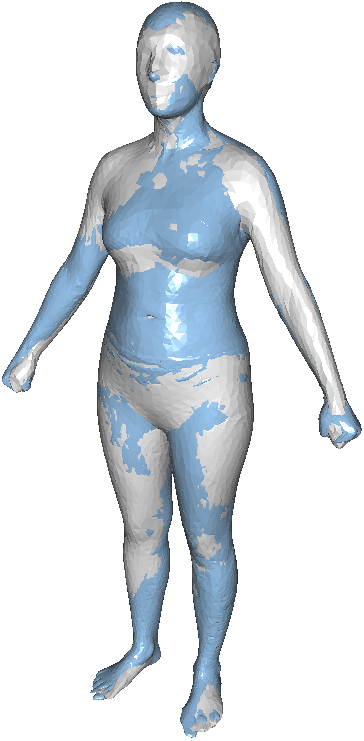}}        
    \hspace{-0.115in}
  \subfigure[]{
    \label{fig:hd0:e} 
    \includegraphics[angle=0,width=0.285in]{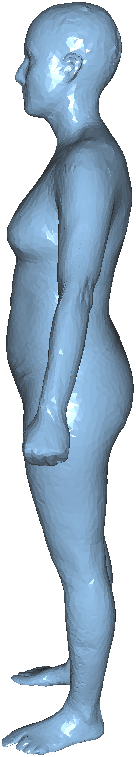}}    
    \hspace{-0.075in}    
  \subfigure[]{
    \label{fig:hd0:f} 
    \includegraphics[angle=0,width=0.275in]{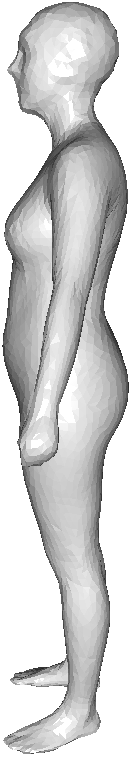}}              
  \caption{\label{fig:hd0}
Example result of our reconstruction from masks of a real-world body with high level details. 
We intend to demonstrate that our body space built on medium level geometry can be used to fit the real-world data well.
  (a) Input 3D human model; (b) masks of the input; (c) result of our method; (d) overlap of (a) and (c);
  (e,f) the lateral views of (a) and (c).
}
\end{figure}

\begin{figure}
  \centering
  \hspace{-0.218in}   
  \subfigure[]{
    \label{fig:hd1:a} 
    \includegraphics[angle=0,width=0.788in]{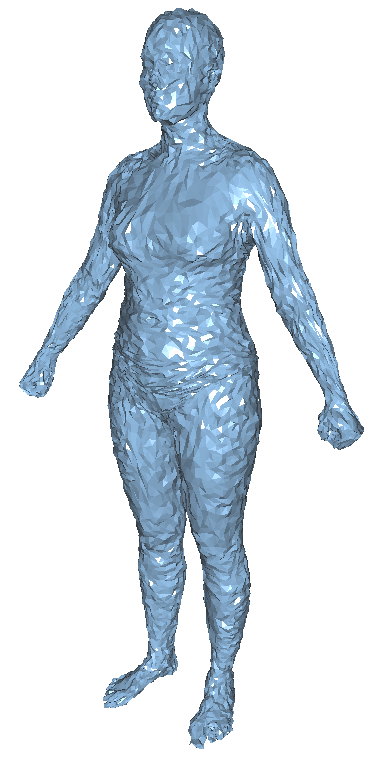}}
    \hspace{-0.115in}
  \subfigure[]{
    \label{fig:hd1:b} 
    \includegraphics[angle=0,width=0.43in]{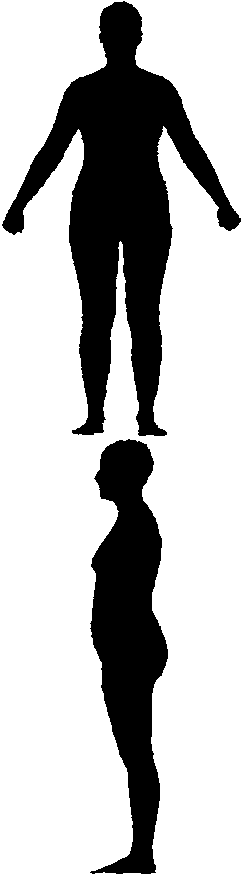}}    
    \hspace{-0.115in}
  \subfigure[]{
    \label{fig:hd1:c} 
    \includegraphics[angle=0,width=0.788in]{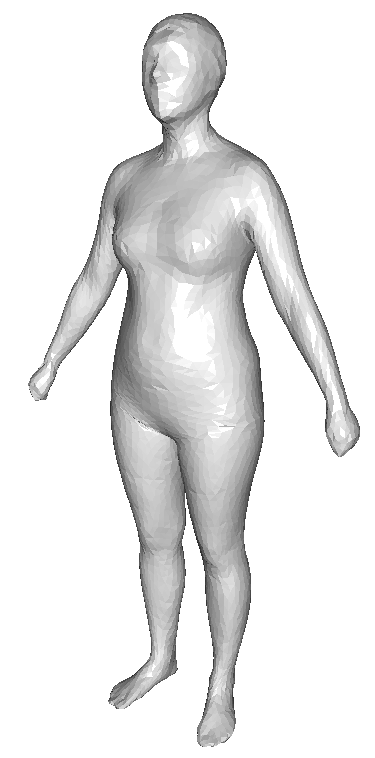}}
    \hspace{-0.135in}
  \subfigure[]{
    \label{fig:hd1:d} 
    \includegraphics[angle=0,width=0.788in]{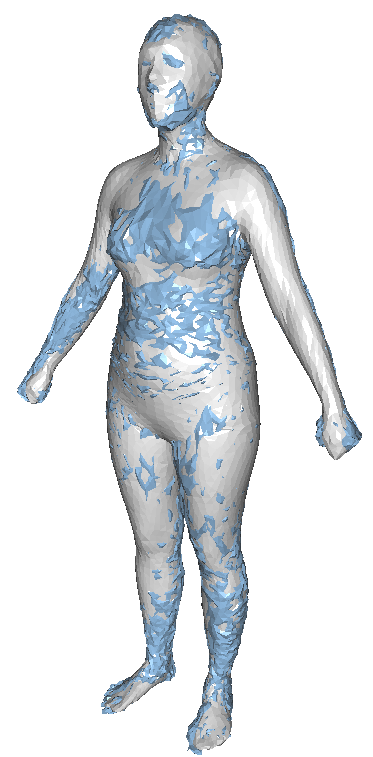}}        
    \hspace{-0.1in}
  \subfigure[]{
    \label{fig:hd1:e} 
    \includegraphics[angle=0,width=0.775in]{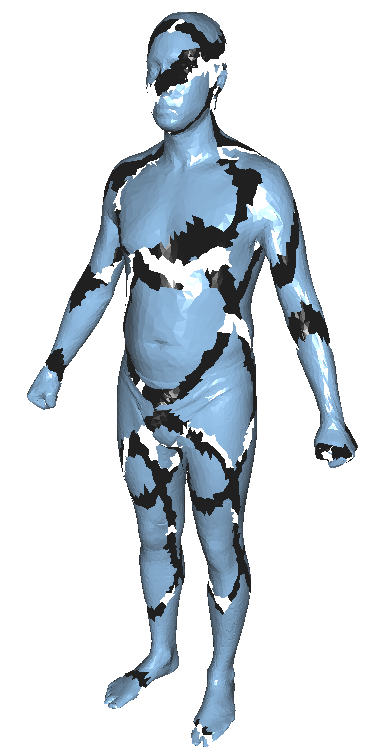}}    
    \hspace{-0.25in}    
  \subfigure[]{
    \label{fig:hd1:f} 
    \includegraphics[angle=0,width=0.72in]{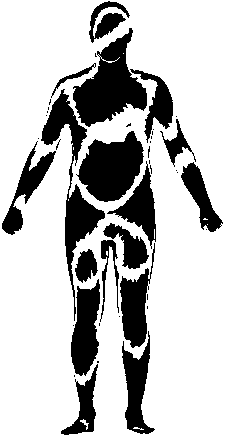}}   
    \hspace{-0.095in}    
  \subfigure[]{
    \label{fig:hd1:g} 
    \includegraphics[angle=0,width=0.25in]{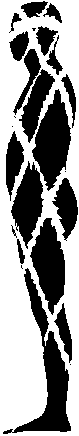}} 
    \hspace{-0.095in}    
  \subfigure[]{
    \label{fig:hd1:j} 
    \includegraphics[angle=0,width=0.71in]{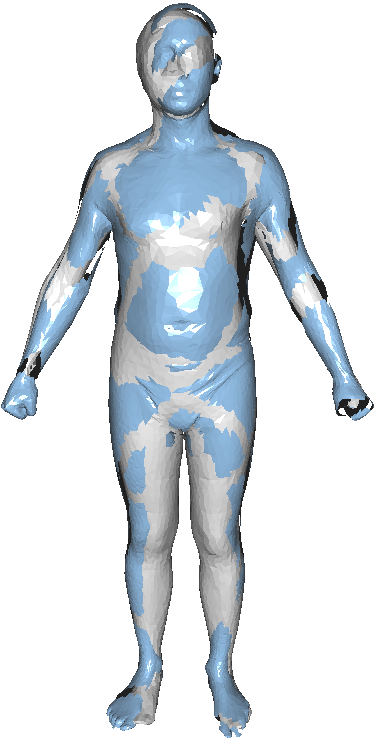}}     
    \hspace{-0.095in}    
  \subfigure[]{
    \label{fig:hd1:h} 
    \includegraphics[angle=0,width=0.72in]{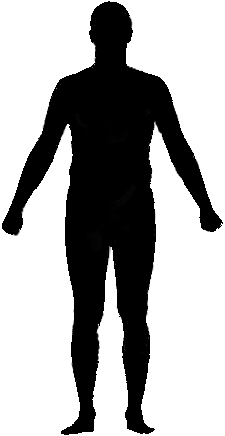}} 
    \hspace{-0.095in}    
  \subfigure[]{
    \label{fig:hd1:i} 
    \includegraphics[angle=0,width=0.25in]{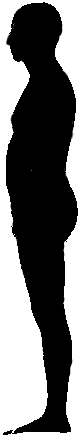}} 
    \hspace{-0.095in}    
  \subfigure[]{
    \label{fig:hd1:k} 
    \includegraphics[angle=0,width=0.71in]{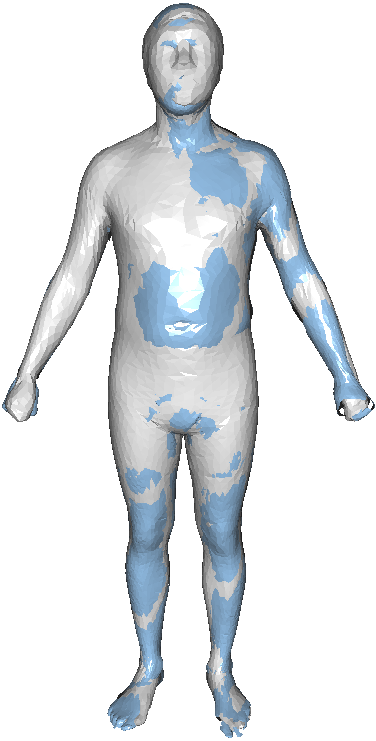}}                             
  \caption{\label{fig:hd1}
Experiments on masks from a noisy model and on masks with holes. 
We intend to demonstrates the generalization of our method on noisy masks.
Thanks to the filters of CNN and the regularization introduced by the body shape space, our method is robust to noisy inputs. The incomplete model is used to demonstrate that our method can be used to
repair body models in the image space, which reduces the effort in obtaining body models from range scans.
  (a) A noisy body model; (b) masks of (a); (c) the resulting body shape; (d) overlap of (a) and (c);
  (e) a incomplete model; (f,g) masks of (e); (h) the resulting body shape of incomplete masks; (i,j) the complete masks; (k)the overlap of (d) and (e).
}
\end{figure}

\textbf{Testing on high-resolution models.}
We also evaluate our reconstruction method on some human bodies with high level details.
Some examples of our reconstruction results are shown in Figure \ref{fig:hd0} and \ref{fig:hd1}.
Our JointMaskNet estimates a body shape within the built body shape space with the masks of a high-resolution body model with 25000 triangles.
Except for the fine scale details and the fists, the reconstructed body of our method fits the original model well. 
In fact, the shapes of fists over our entire body space are distorted due to the resolution.
%%% Figure
%%%

To evaluate the performance of our method on the noisy masks, 
we add heavy Gaussian noise to a high-resolution body model, 
and then apply our JointMaskNet on the noisy masks.
As can be seen from Figure \ref{fig:hd0:c} and Figure \ref{fig:hd1:c}, the difference between two results is quite small,
which demonstrates that our method generalizes well on noisy masks and can produce robust results.
To mimic incomplete range scan with holes, we remove some triangle strips from a body model (see Figure \ref{fig:hd1:e}).
The mask images of the incomplete model are shown in Figure \ref{fig:hd1:f} and Figure \ref{fig:hd1:g}.
When we put them into the JointMaskNet directly, we obtain a body which is slightly thinner than the original one as shown in Figure \ref{fig:hd1:h}.
A defective mask image seems to our JointMaskNet that it is an attenuated signal coming from a thinner body.
Then, we manually restore the mask image complete with a 2D brush in image space,
obtaining a more reasonable result.
These examples also imply that our method can be used to generate body models fitting range scans without any marker data, with a little effort in 2D image space.
The resulting models share the same topology, without substantial effort to process the noisy and incomplete surface in 3D space.
%%% Figure
%%%

\begin{figure}
  \centering
  \hspace{-0.138in}  
  \subfigure{
    \label{fig:skeleton:a} 
    \includegraphics[angle=0,width=0.65in]{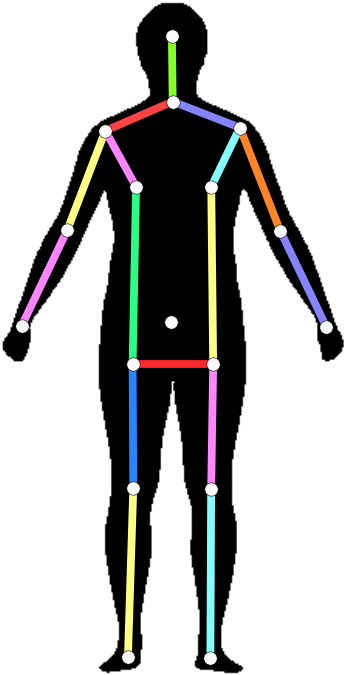}}
  \hspace{-0.09in}
  \subfigure{
    \label{fig:skeleton:b} 
    \includegraphics[angle=0,width=0.705in]{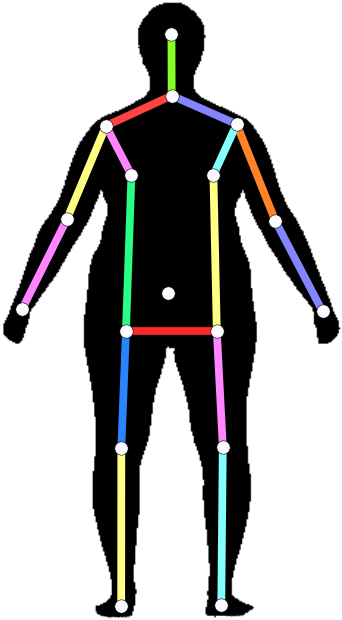}}
    \hspace{-0.09in}
  \subfigure{
    \label{fig:skeleton:c} 
    \includegraphics[angle=0,width=0.62in]{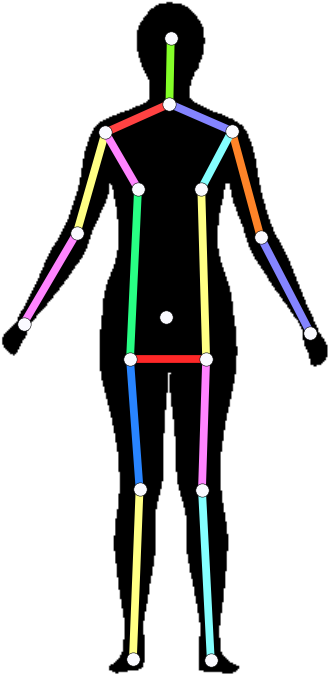}}
    \hspace{-0.09in}
  \subfigure{
    \label{fig:skeleton:d} 
    \includegraphics[angle=0,width=0.76in]{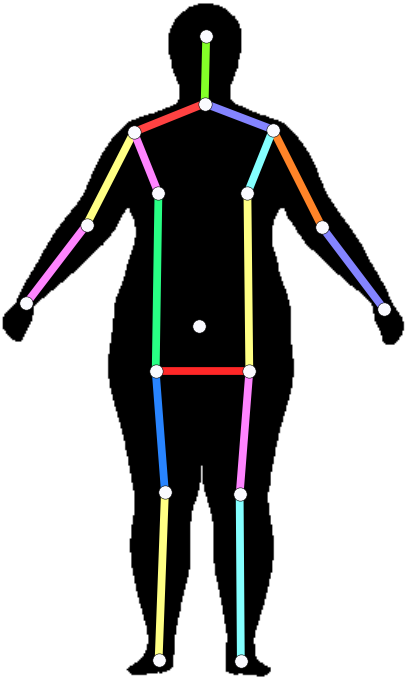}}
    \hspace{-0.09in}
  \subfigure{
    \label{fig:skeleton:e} 
    \includegraphics[angle=0,width=0.725in]{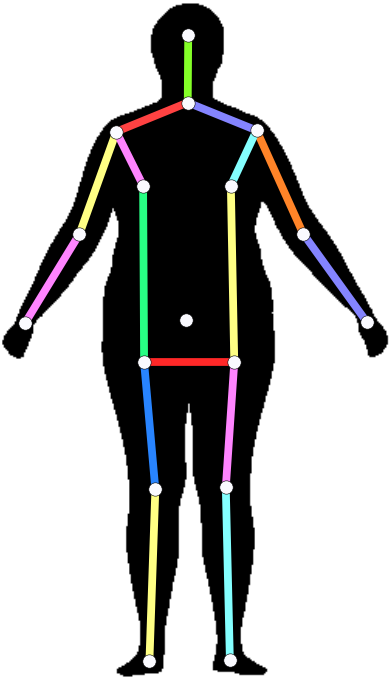}}        
  \caption{\label{fig:skeleton}
Skeletons extracted from the 3D body shapes generated by JointMaskNet. 
}
\end{figure}

\textbf{Skeleton estimation.}
The proposed method estimates the body parameters directly from binary images without relying on extra 2D annotations, 
such as segmentation or joint keypoints.
However, we can extract these feature information afterwards from the 3D meshes generated by our method. 
Figure \ref{fig:skeleton} presents some extracted skeletons.
We train a joint regressor to predict the joint postion from the known geometry of body, 
then use it to estimate the joints from the newly generated body shapes, 
and finally we project the 3D locations to the image plane.
We can see that the joints are located at appropriate positions,
which also implies that our CNN-based network predicts the body parameters accurately. 
%%% Figure
%%%

\textbf{Limitations and discussion.}
Although our method is capable of recovering a wide range of body shapes from masks due to our use of such a large dataset to train a dense CNN-based network,
there are still some types of shapes out of the built body space. 
We can address this issue by expanding the training dataset further.
Our current method focuses on constructing human body shapes in a simple and easy manner,
consequently, it generates mesh models with medium scale geometry due to the absence of fine scale details in binary mask images.
Anyhow, the result of our networks can be used as an initialization
and anchor for the further optimization.
For example, users might improve the shape in an interactive way, such as by modifying the local shapes with the interactive user aid indicated via sketching \cite{SketchBasedEditing2005}. 
And fine scale geometry details can be extracted using exist shape from shading methods, and
the reconstructed shape model can be improved by adding them directly or by encoding them in a normal map or a displacement map.
Especially, details on face can be enhanced by recently developed CNN-based methods which reconstructing fine scale details from RGB images \cite{RichardsonSK16,tran2016regressing,Guo20183DFace}.
In this paper, we focus on developing a 3D modeling algorithm for translating 2D profiles in orthographic views into a 3D body shape which is similar to the traditional 3D modeling systems \cite{Rivers2010}.
In this paper, the masks are given in advance.
For example, they can be captured from an existing 3D body as shown in Figure\ref {fig:hd0} and \ref{fig:hd1}.
And our algorithm can be used to estimate a base surface, to fill holes and to extract the skeleton for a given human body model.
Alternatively, starting from template masks, users can edit them to match a body in an image.
The estimation for the accurate masks and the camera parameters from a RGB image is not our current target in this paper.
In addition, to estimate the shape of arms and legs more accurately, 
our method limits the body shapes in standard posture, and it might fail to capture large poses. 
In fact, the built shape space, with 50 principal components, has already encoded a small range of poses existing in the original dataset, thus it allows small changes in pose (see the head, arms and legs in
Figure \ref{fig:singleview0}, \ref{fig:doubleview1} and \ref{fig:skeleton}).

\subsection{Conclusions and future work}

Automatic reconstruction human body from color images in clothing might impact the accuracy of geometry
due to the ambiguity introduced by self-occlusions, 
while digital nude photographs may cause the privacy issues.
To overcome these limitations, we use binary mask images in this paper. 
We design a novel CNN-based regression network with two branches to estimating 3D human body shape from 2D mask images.
In addition, to overcome the shortage of training data required for this purpose, 
we propose some significantly data augmentation schemes for 3D human bodies, 
which can be to promote further research on this topic.
Experiments demonstrate the effectiveness of our approach to produce accurate results.
Our method make it easy to create digital human body for 3D game, virtual reality and online fashion shopping. 
Further more, our method could be easily and directly extended to reconstruct high-resolution human body if a wide range of bodies with fine scale details had been acquired. 

In current implementation, our method needs two mask images to accurately predict the body parameters.
An alternative scheme is that given a single mask image, we can train a network to predict each other.
In addition, it deserves further study
to evaluate an accurate mask directly from a single RGB image automatically or in an easy interactive way. We leave these features to the future work. 

% Bibliography
\bibliographystyle{ACM-Reference-Format}
\bibliography{humanbody2018}

\end{document}